\documentclass[12pt]{article}

\usepackage{amssymb,amsmath,amsfonts,array,eurosym,geometry,graphicx,color,setspace,sectsty,comment,footmisc,caption,natbib,pdflscape,float,booktabs,appendix,tabularx,mathtools,threeparttable,rotating,lscape,longtable,enumerate,bbm,multirow,enumitem,subcaption,arydshln,tikz,dcolumn}

\usepackage[export]{adjustbox}
\usetikzlibrary{positioning, decorations.markings,decorations.pathreplacing,calligraphy}
\usepackage{hyperref}
\hypersetup{
colorlinks=true,
citecolor=blue,
linkcolor=blue,
filecolor=blue,      
urlcolor=blue,
}

\def\sym#1{\ifmmode^{#1}\else\(^{#1}\)\fi}
\begin{document}

\begin{titlepage}
\singlespacing
\title{\Large Learning by exporting with a dose-response function \thanks{\scriptsize{We would like to thank Gabor Bekes, Giovanna D'Inverno, Daniela Maggiorni, Marianna Marino for useful comments. We are also grateful to participants at the European Trade Study Group (ETSG) in 2024 in Athens. Armando Rungi claims financial support from the PRIN CUP: D53D23006740006, funded by the European Union – NextGenerationEU.}}}

\author{Francesca Micocci\thanks{\scriptsize Mail to: francesca.micocci@uniroma3.it.  Department of Economics,  Roma Tre University,  Via Silvio D'Amico 77-00145 Rome,  Italy.} \and Armando Rungi\thanks{\scriptsize Mail to: armando.rungi@imtlucca.it. Laboratory for the Analysis of Complex Economic Systems, IMT School for Advanced Studies, piazza San Francesco 19 - 55100 Lucca, Italy.} \and
Giovanni Cerulli\thanks{\scriptsize Mail to: giovanni.cerulli@ircres.cnr.it. IRCRES-CNR, Research Institute on Sustainable Economic Growth}}

\date{\small{This version: April 2025} }
\maketitle
\vspace{-2.5em}

\begin{abstract}
\footnotesize
\singlespacing
\noindent  
This paper investigates the causal effect of export intensity on productivity and other firm-level outcomes with a dose-response function.  After positing that export intensity acts as a continuous treatment, we investigate counterfactual productivity levels in a quasi-experimental setting.  For our purpose, we exploit a control group of non-temporary exporters that have already sustained the fixed costs of reaching foreign markets, thus controlling for self-selection into exporting. Our findings reveal a non-linear relationship between export intensity and productivity, with small albeit statistically significant benefits ranging from 0.1\% to 0.6\% \textit{per year} only after exports reach 60\% of total revenues. After we look at sales, variable costs, capital intensity, and the propensity to filing patents, we show that, before the 60\% threshold, economies of scale and capital adjustment offset each other and induce, on average, a minimal albeit statistically significant loss in productivity of about 0.01\% \textit{per year}. Crucially, we find that heterogeneous export intensity is associated with the firm's position on the technological frontier, as the propensity to file a patent increases when export intensity ranges in 8\%-60\% with a peak at 40\%. The latest finding further highlights that learning-by-exporting is linked to the building of absorptive capacity.

\vspace{20pt}

\bigskip
\end{abstract}
\setcounter{page}{0}
\thispagestyle{empty}
\end{titlepage}
\pagebreak \newpage

\onehalfspacing
\section{Introduction}\label{c3: intro} 

Previous literature has extensively studied the relationship between productivity and exporting. The main challenge was to unravel reverse causality and check which mechanism prevails. On the one hand, there is a self-selection mechanism into the exporting status, by which only the most productive firms can reach foreign markets because beachhead costs are relevant \citep{Roberts_Tybout_1997, bernard1999exceptional, melitz2003impact, bernard2007firms, Melitz&Ottaviano2008, Bernardetal2012}. On the other hand, there is a mechanism of learning by exporting (LBE), by which a firm's productivity improves after entering a foreign market thanks to knowledge spillovers coming directly from buyers or through increased competition from foreign producers \citep{Sofronis_et_al_1998, baldwin2003export, Crespi_et_al_2008, DeLoecker_2013, Atkin_et_al_2017, Liang_et_al_2024}. 

Our perspective is different. Our aim is to investigate the effect of export intensity on a firm's productivity by adopting a potential outcome framework \citet{imbens2010rubin} and, thus, estimate a dose-response function once we assume that export intensity represents a dose of a treatment \citep{hirano2004propensity, kluve2012evaluating, bia2014stata, cerulli2015ctreatreg, d2023nonparametric}. Briefly, we test whether firms react heterogeneously to different levels of export intensity after they have already decided to export, therefore, after the self-selection mechanism into exporting status has already manifested. With a quasi-experimental approach, we administer doses of exporting to these firms and check the effect of the treatment after comparing them with counterfactual exporters that did not receive that dose.

Our central intuition is that firms' productivity benefits from exporting only after reaching a certain capacity level. When export intensity is low, firms must first establish the necessary absorptive capacity - i.e., their ability to utilize external knowledge - and create an effective logistical organization to fully realize productivity gains from foreign markets. 

Our hypothesis is confirmed after investigating French exporters from 2010-2018. In particular, we estimate a dose-response curve following \citet{cerulli2015ctreatreg}, where the Average Treatment Effect on Treated (ATET) is conditional upon the observable characteristics of firms, with firm-level fixed effects accounting for unobserved, time-invariant differences. Crucially, our control group is made up of non-temporary exporters - firms that have already sustained the sunk costs of reaching foreign markets. Excluding both non-exporters and temporary exporters from our control group allows us to identify the LBE productivity gains better and separate them from the productivity premia that are needed to self-select into the exporting status. Our baseline application presumes a third-degree polynomial to represent the continuous impact of export intensity on firm performance. We first test the impact on firm-level Total Factor Productivity (TFP), estimated following \cite{ackerberg2015identification}, and then we consider alternative outcomes (sales, variable costs, capital intensity, propensity to file a patent) to investigate the channels through which TFP benefits can arise.

As expected, the typical dose-response curve shows that the relationship between TFP and export intensity is non-linear. We find that, on average, the benefits from LBE are statistically significant only after an export intensity equal to 60\%. Our estimates indicate a modest TFP rise ranging from 0.1\% to 0.6\% \textit{per year}. On the other hand, the non-linear shape of the dose-response curve points to a segment where minor albeit statistically significant productivity losses are recorded (0.01\%) when firms have export intensity between 5\% and 35\% of export intensity. When we explore costs, capital intensity, and patent filings, we find that the latter is a segment where economies of scale are built and, at the same time, capital intensity increases and innovations are made. We argue that, as firms expand their export activities, they seek to remain competitive in foreign markets by investing in cutting-edge technologies and organizational capabilities. However, adjusting capital structures and investing in research and development to meet these demands can be costly and may temporarily depress productivity during the transition.

The rest of the paper is structured as follows. Section \ref{sec: data} introduces the data, while in Section \ref{sec: methods}, we outline our estimation strategy. Results are discussed in Section \ref{sec: results}. We then present some robustness and sensitivity checks in Section \ref{sec: robustness}. Finally, we discuss limitations in Section \ref{sec: limitations} and sketch conclusions in Section \ref{sec: conclusion}.

\section{Data and descriptive statistics}
\label{sec: data}
We source firm-level information for French exporters in the time interval 2010-2018 from Orbis, by Moody's\footnote{The Orbis database by Moody's is a recognized global source for firm-level financial accounts and has been extensively used in previous studies, including \cite{gopinath2017capital}, \cite{cravinolevchenko}, \cite{DelPrete&Rungi2017}, \cite{RungiDelPrete2018}, and \citet{Micocci_Rungi_2023}.}. In particular, we focus on France as it is a well-explored case study for firm-level trade data, providing a foundation for building upon and confronting previous literature. See, among others, \cite{crozet2012quality} and \citet{FontagneSecchiTomasi}.

Our primary variables of interest are a firm's export intensity, which we derive from information about export revenues\footnote{French firms must report revenues from exports separately, as from the subsequently amended \textit{Règlement n. 99-03 du Comité de la réglementation comptable. It makes the French case peculiar in the Orbis database, as it is the only case for which we have systematic reliable information on exports.}} out of the total revenues, and Total Factor Productivity (TFP), which we estimate from a production function following \citet{ackerberg2015identification}. To assemble the production function, we proxy capital with firm-level fixed assets, labor with the number of employees, and intermediate inputs with the cost of materials. Nominal values are deflated using producer price indices made available by the national statistics offices (INSEE). When we discuss channels through which export intensity can have an impact on TFP, we use additional firm-level outcomes, including variable costs (costs of employees and costs of intermediate inputs), capital intensity (ratio of fixed assets on employees), and the portfolio of registered patents. See Appendix Table \ref{app_tab: variables} for more details on firm-level accounts.

For our sample, we select firms active in manufacturing industries as they are the ones delivering goods that cross national borders. Mainly, we want to exclude from our sample the case of trade intermediaries, who professionally receive goods and sell them to foreign markets on behalf of other (manufacturing) firms. We argue that the LBE mechanism would not apply to trade intermediaries. After estimating TFP, we consider only firms that have engaged at least once in exporting in our analysis period. This is important to rule out the endogeneity between productivity and export status. Finally, to avoid noise in the relationship between export intensity and productivity, our baseline models consider non-temporary exporters, i.e., those firms that export for at least four consecutive years \citep{BekesMurakozy} in our analyses. Please, see Section \ref{sec: methods} for a detailed discussion of our identification strategy. 

\begin{figure}
\caption{Our sample and the population of exporters in France}
\label{fig: coverage}
    \centering
\includegraphics[width=.5\textwidth]{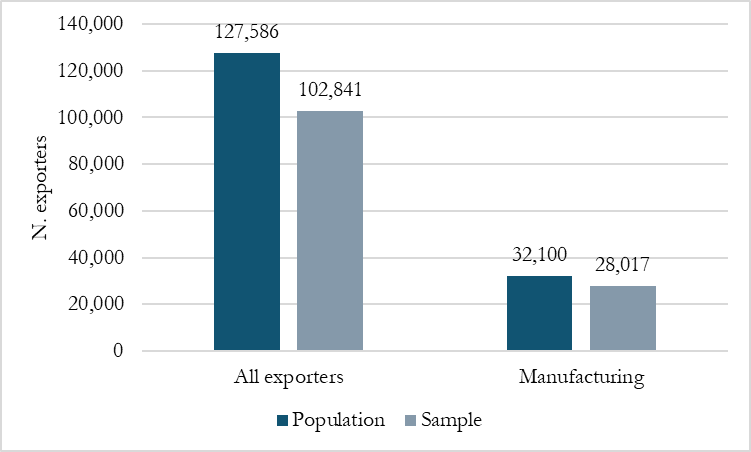}
    \begin{tablenotes}
\singlespacing
\footnotesize
    \item Note: The Figure reports sample coverage of French exporters from the Orbis database and population numbers from \citet{INSEE}. We investigate manufacturing exporters to avoid the inclusion of trade intermediaries and other services firms that do not professionally deliver goods to foreign markets. 
\end{tablenotes}
\end{figure}

\begin{figure}[!tbp]
  \centering
  \begin{minipage}[b]{0.35\textwidth}
    \includegraphics[width=\textwidth]{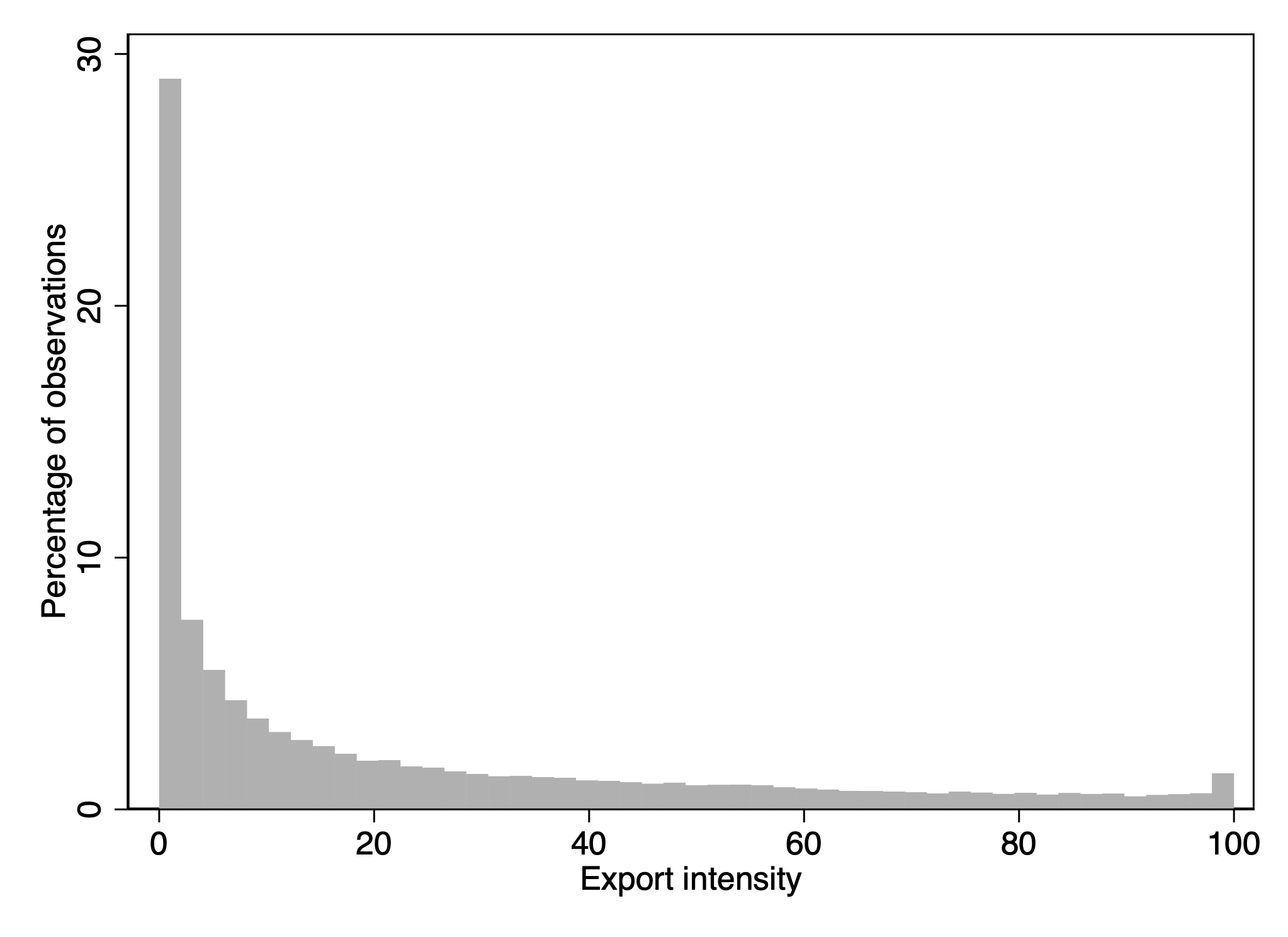}
    \caption{Export intensity}
    \label{fig: export intensity}
  \end{minipage}
  \hfill
  \begin{minipage}[b]{0.5\textwidth}
    \includegraphics[width=\textwidth]{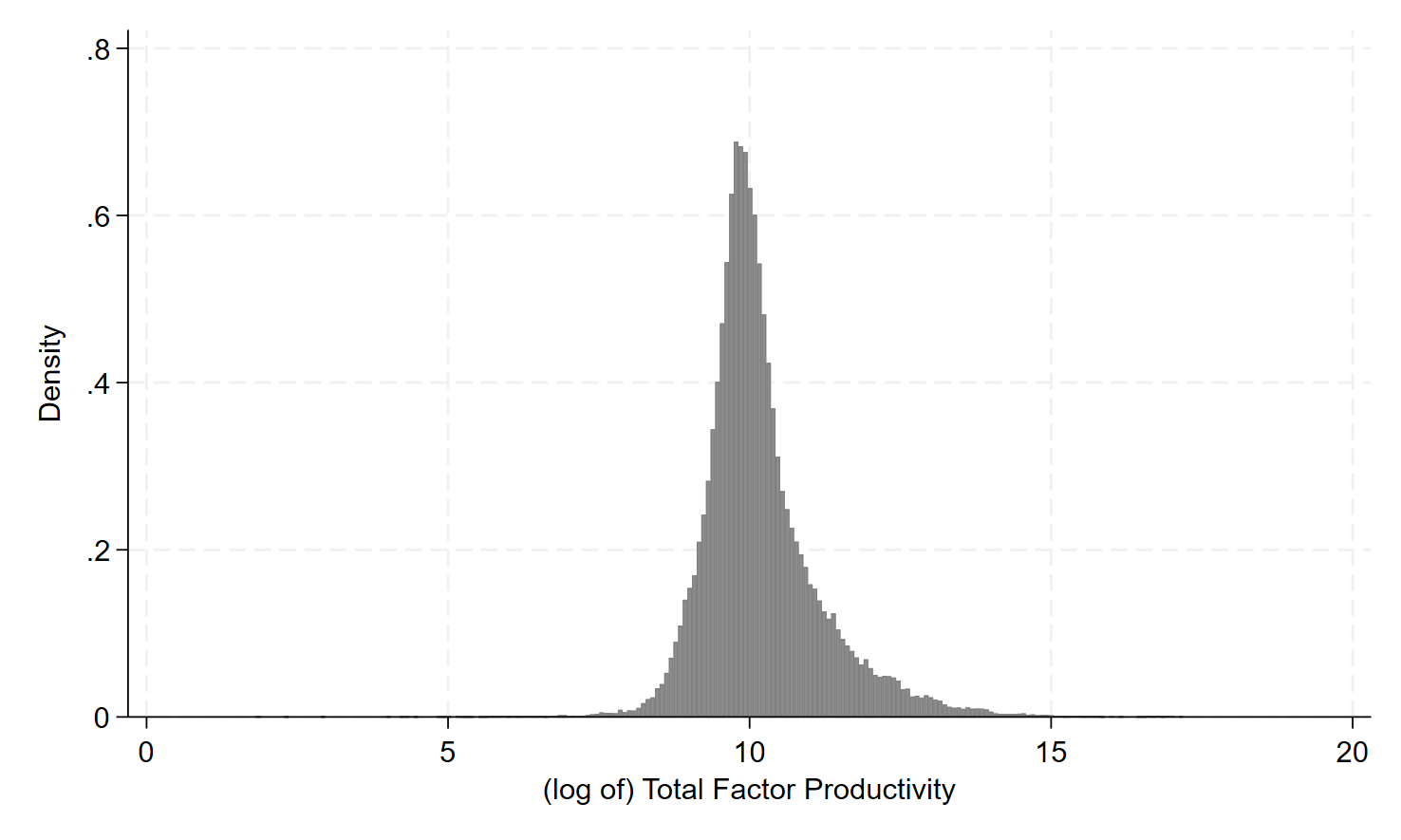}
    \caption{(log of) TFP distribution}
    \label{fig: tfp distribution}
  \end{minipage}
      \begin{tablenotes}
\singlespacing
\footnotesize
    \item Note: On the left, Figure \ref{fig: export intensity} shows the distribution of firm-year observations of export intensity by French exporters in the period 2010-2018. On the right, Figure \ref{fig: tfp distribution} shows the distribution of firm-year observations in terms of TFP in the period 2010-2018 estimated following the production function approach by \cite{ackerberg2015identification}.
\end{tablenotes}
\end{figure}

In Figure \ref{fig: coverage}, we report a snapshot of our sample coverage. If we compare with national statistics offices, we have 87\% of exporters out of the total of manufacturing exporters in the population of firms. Figures \ref{fig: export intensity} and \ref{fig: tfp distribution} focus on non-temporary exporters\footnote{Please note that descriptive evidence and baseline analyses are based on the subsample of non-temporary exporters, whereas the description of the total sample includes temporary exporters as they are used for a few robustness checks in Section \ref{sec: robustness}.} and show how our main variables of interest are distributed. We compute export intensity as a simple ratio between export and total revenues, thus bounded in a range $[0. 1]$. Notably, we observe that the distribution of export intensity has a long right tail because most firms export less than 10\%. Only 1.3\% exporters reach an intensity of $100\%$, selling exclusively abroad. Let's move our attention to the TFP distribution in Figure \ref{fig: tfp distribution}. We know that it is much more skewed than export intensity, and we had to transform it in logs to visualize the distribution. In line with previous literature, we have that exporters have, on average, a higher TFP. See Appendix Table \ref{tab: export_premium_stats}. When we focus on exporters, as in Figure \ref{fig: tfp distribution}, we find non-negligible evidence of firm-level heterogeneity. Even in logs and after excluding the subsample of non-exporters, the TFPS show a skewness to the right, where a fringe of the most efficient exporters is. In the rest of the paper, we want to explain how much of this heterogeneity can be explained by a different exposure to export intensity.

\begin{figure}[htpb!]
\caption{Export intensity and TFP}
\centering
  \includegraphics[width=0.5\textwidth]{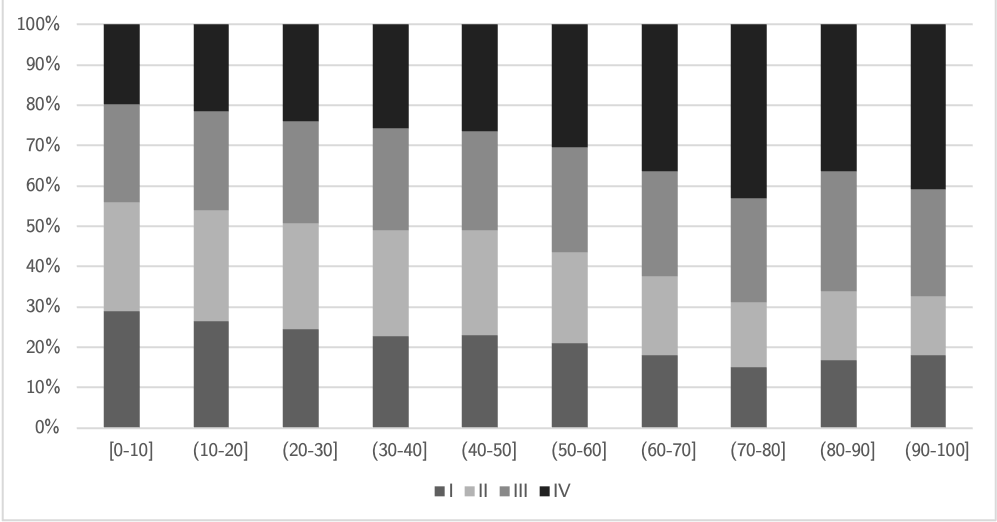}
  \label{fig:export_intensity_TFP}
\begin{tablenotes}
    \singlespacing \footnotesize
    \item Note: The figure focuses on non-temporary exporters to show export intensity intervals on the x-axis and TFP distribution partitioned by quartiles on the y-axis. We observe that the last TFP quartiles have a bigger support when we move to higher-intensity intervals. Yet, TFP variation is relatively large along the export intensity distribution.
\end{tablenotes}
\end{figure}

Table \ref{tab: intensity and size} and Figure \ref{fig:export_intensity_TFP} investigate joint distributions. In Table \ref{tab: intensity and size}, we have a distribution of export intensity across firm size categories, once considering each firm-year observation present in our sample. A description of the firm size categories is provided in Appendix Table \ref{app_tab: variables}. In Figure \ref{fig:export_intensity_TFP}, we consider once again firm-year observations and we plot TFP quartiles along export intensity intervals.

What we immediately observe is that there is a good share of exporters that have not exported in some years. Please remember that we are looking at the subsample of non-temporary exporters. Although the definition of non-temporary exporters is stringent because it implies that a firm needs to export for at least four consecutive years, we do have years in which those exporters do not sell abroad. A majority of them are present in the medium category (47\%), but we can observe that they can be present across all size categories. On average, export intensity is higher when firm size increases, but we still have a high variation of export intensity in all size categories. Similarly, we observe that TFP is higher when export intensity increases. Yet, we can find that firms of all sizes and TFPs are represented across all levels of export intensity. Notably, what we find is preliminary evidence that is not only relevant \textit{per se}, as it shows that heterogeneity can play in different directions. It is also helpful because a higher variation in these covariates allows us to design our identification strategy better.

\begin{table}[!ht]
\caption{Export intensity and firm size}
    \centering
    \resizebox{.6\textwidth}{!}{%
    \begin{tabular}{cccccc}
    \hline
        Export intensity & Small & Medium & Large & Very large & Total \\ \hline
        0 & 38.30 & 47.01 & 12.61 & 2.00 & 100 \\ 
        (0-10] & 22.77 & 51.18 & 23.13 & 2.92 & 100 \\ 
        (10-20] & 17.71 & 49.33 & 28.46 & 4.50 & 100 \\ 
        (20-30] & 15.73 & 46.90 & 32.04 & 5.33 & 100 \\ 
        (30-40] & 14.04 & 44.82 & 34.79 & 6.35 & 100 \\ 
        (40-50] & 13.13 & 41.15 & 37.06 & 8.67 & 100 \\ 
        (50-60] & 11.97 & 37.80 & 40.81 & 9.42 & 100 \\ 
        (60-70] & 10.40 & 36.15 & 41.40 & 12.05 & 100 \\ 
        (70-80] & 10.71 & 30.50 & 45.32 & 13.47 & 100 \\ 
        (80-90] & 10.27 & 29.82 & 44.11 & 15.80 & 100 \\ 
        (90-100] & 15.88 & 29.33 & 41.12 & 13.67 & 100 \\ \hline
    \end{tabular}}
    \begin{tablenotes}
    \singlespacing \footnotesize
    \item Note: The table focuses on non-temporary exporters as defined by \citet{BekesMurakozy}, and it shows the joint distribution of export intensity and firm size. Non-temporary exporters can still show zero exports in some years, and this is relevant for the construction of a control group for our dose-response exercises.
\end{tablenotes}
    \label{tab: intensity and size}
\end{table}

\newpage

\section{Empirical strategy}
\label{sec: methods}
Our aim is to investigate the causal impact of export intensity on productivity. To do so, we adopt the dose-response function proposed by \cite{cerulli2015ctreatreg}.\footnote{For a review of the recent literature on dose-response models, see \cite{hirano2004propensity,kluve2012evaluating,bia2014stata,d2023nonparametric}.} Dose-response models are particularly well-suited for contexts like ours, where we are not interested in the binary treatment (exporter vs. non-exporter), but we focus on the degree of exposure (i.e., the 'dose' of exports) experienced by the treated company. A useful, distinctive feature of the dose-response approach is that we can examine an entire distribution of causal effects while presenting results in an intuitive graphical format.

Let us start with Rubin's potential outcome equation:
\begin{equation}
\label{eq: Rubin}
    y_i=y_{0,i}+w_i(y_{1,i}-y_{0,i})
\end{equation}
where $y_{0, i}$ represents the potential productivity for company $i$ if untreated, while $y_{1, i}$ is the potential productivity when treated, and $w_{i}$  is a dummy variable indicating treatment status. In our framework, the treatment consists of having positive export revenues in $t-1$.

Expanding this equation into a continuous framework, we introduce export intensity, $d_i$, as a continuous treatment indicator ranging in $[0, 100]$, computed as the ratio between export revenues and total revenues for the $i$-th firm in percentage points. Therefore, we can model the relationship between export intensity and productivity with $h(d_i)$, a general differentiable function, and we can introduce $g(x_i)$ as a function of the firm's characteristics that can be considered as confounders, $\textbf{x}_i = [x_{1,i}, x_{2,i}, \dots , x_{M,i}]$. Notably, $\mu_1$ and $\mu_0$ are scalars, and $e_1$ and $e_0$ are error terms corresponding to random variables with an unconditional mean equal to zero and constant variance. The population-generating process for the two potential outcomes can be thus expressed as follows:
\begin{equation}
    \begin{cases}
    \label{eq: pot_outcomes}
        w=1: \quad y_1=\mu_1+g(\mathbf{x})+h(d)+e_1\\
        w=0: \quad y_0=\mu_0+g(\mathbf{x})+e_0
    \end{cases}
\end{equation}
where we get rid of indexes $i$ and $t$ for the sake of conciseness, and where the function $h(d)$ is non-zero only when a company is in the treated status, $w=1$. 

In our context, the control group is made of companies that have already been able to reach foreign markets. The latter condition allows us to focus on firms that have already sustained the fixed costs of exporting, thus self-selecting into the export status thanks to their relatively higher productivity if compared with non-exporters \citet{Roberts_Tybout_1997, bernard1999exceptional, melitz2003impact, Bernardetal2012}. Among exporters, we specifically focus on the category of non-temporary exporters \citet{BekesMurakozy} to avoid noise in the relationship with productivity. The minimum condition for a non-temporary exporter is that it exports at least four consecutive years in our period of analyses. Crucially, as observed in Table \ref{tab: intensity and size}, there is still enough variation in our sample that we can use to build our control group when $w=0$.

At this point, we can define the treatment effect as $TE = (y_1 - y_0)$, and we obtain our causal parameters of interest, i.e.,  the Average Treatment Effect of export intensity conditional on firms' observable characteristics, $\mathbf{x}$, at different levels of treatment, $d$:
\begin{equation}
    ATE(\mathbf{x}, d ) = E(y_1 - y_0|\mathbf{x}, d)
\end{equation}
By the law of iterated expectation, the corresponding population unconditional ATE can be obtained as:
\begin{equation}
    ATE=E_{(\mathbf{x},d)}\{ATE(\mathbf{x},d)\}
\end{equation}
If we assume a linear-in-parameters form for $g(\mathbf{x}) = \mathbf{x\delta}$, the \textit{ATE} conditional on $\mathbf{x}$, $d$, and $w$ becomes:
\begin{equation}
\label{eq: ATE_def}
    ATE(\mathbf{x}, d, w) = w \times \{\mu + h(d)\} + (1 - w) \times  \{\mu\}
\end{equation}
where $\mu=(\mu_1-\mu_0)$. The corresponding unconditional \textit{ATE} will be:
\begin{equation}
\label{eq: uncond_ATE}
    ATE=p(w=1)\times (\mu+\overline{h}_{d>0})+p(w=0)\times (\mu)
\end{equation}
where $p(w=1)$ is the probability of being an exporter, and $h_{d>0}$ is the average of the response function taken over $d>0$, i.e., the average productivity gains from export intensity.

Substituting the potential outcomes in Eq. (\ref{eq: pot_outcomes}) into Rubin’s potential outcome Eq. (\ref{eq: Rubin}), we obtain:
\begin{align*}
    y&=y_0+w(y_1-y_0)\\    
&=\mu_0+\mathbf{x}\delta+\epsilon_0+w[(\mu_1+\mathbf{x}\delta+h(d)+\epsilon_1)-(\mu_0+\mathbf{x}\delta+\epsilon_0)]\\
&=\mu_0+\mathbf{x}\delta+w(\mu_1-\mu_0)+w(h(d))+\epsilon_0+w(\epsilon_1-\epsilon_0)\mathbf{+w\overline{h}-w\overline{h}}\\
&=\mu_0+\mathbf{x}\delta+w(\mu_1-\mu_0+\overline{h})+w(h(d)-\overline{h})+\epsilon_0+w(\epsilon_1-\epsilon_0)\\
&=\mu_0+\mathbf{x}\delta+wATE+w(h(d)-\overline{h})+\eta
\end{align*}

The previous expression can be estimated assuming a third-degree polynomial for $h(d_i)$ and firm-level fixed-effects:
    \begin{equation}
    \label{eq: regression}
\ddot{y}_{it}=\alpha_0+\ddot{\mathbf{x}}_{it}\delta_0+w_{it}ATE+w_{it}[a\ddot{D}_{1it}+b\ddot{D}_{2it}+c\ddot{D}_{3it}]+\ddot{\eta}_i
\end{equation}
where we compute the within-variation for each variable $\ddot{v}$ as $v_{it}-\overline{v}_i+\overline{\overline{v}}$, i.e., the deviation from the individual mean of the period $\overline{v}_i=\sum_t v_{it}/t$, plus the population mean of variable $v$ for the whole period. The constant $\alpha_0$ is the average value of the fixed effects, i.e., the grand average of $y$ across all units and all periods. Finally, the polynomial terms are included as $D_j=d^j-E(d^j)$ for $j=1,2,3$. 

Assuming conditional mean independence, a least-squares estimation of Eq. (\ref{eq: regression}) produces consistent estimates of the parameters $\hat{\delta}_0$, $\hat{ATE}$, $\hat{a}$, $\hat{b}$, and $\hat{c}$. With these parameters at hand, we can finally estimate the dose-response function as: 
\begin{equation}
\label{eq: drf}
\begin{split}
    \hat{ATE}(\ddot{t}_{it})=&w\left[\hat{ATE}_{d>0}+\hat{a}\left(\ddot{d}_{it}-\frac{1}{NT}\sum_{i=1}^N\sum_{i=1}^D \ddot{d}_{it}\right)+\hat{b}\left(\ddot{d}^2_{it}-\frac{1}{NT}\sum_{i=1}^N\sum_{i=1}^D \ddot{d}^2_{it}\right)\right.\\
    &\left.+\hat{c}\left(\ddot{d}^3_{it}-\frac{1}{NT}\sum_{i=1}^N\sum_{i=1}^D \ddot{t}^3_{it}\right) \right] +(1-w)\hat{ATE}_{t=0}
    \end{split}
\end{equation}

where the main outcome of interest is the Total Factor Productivity (TFP) for the $i$-th firm at time $t$, which we estimate following \citet{ackerberg2015identification}. In the following paragraphs, we will further investigate alternative outcomes (sales, variable costs, capital intensity) to unravel the channels through which export intensity affects TFP. Please note how firm-level fixed effects account for a substantial portion of the time-invariant heterogeneity. Nonetheless, we include a set of time-varying covariates, $X_{it}$, to address residual time-varying endogeneity. Specifically, we control: a) for changes in firm size with (the logarithm of) the number of employees; b) for changing financial constraints with the size-age indicator proposed by \citet{hadlock2010new}; for variation in innovation dynamics by counting patents in a firm's portfolio. Please note that the validity of our identification strategy hinges on the assumption that there is no omitted firm-level characteristic that simultaneously influences both the outcome $y$ and the continuous treatment $d$ over time. We will challenge this assumption in Section \ref{sec: robustness}. 

\section{Results}
\label{sec: results}
In the following paragraphs, we first introduce our results on the causal relationship between TFP and export intensity. Then we investigate the relationship with export intensity of alternative firm-level outcomes, including sales, costs, capital intensity, and the propensity to file a patent. We aim first to show the heterogeneous impact of export intensity on productivity and then infer which channels determine productivity gains or losses.

\subsection{The impact of export intensity on productivity}

\begin{figure}[htpb!]
    \centering
        \caption{Dose-response function of export intensity on TFP}
    \label{fig:drf_TFP}
    \includegraphics[width=0.6\textwidth]{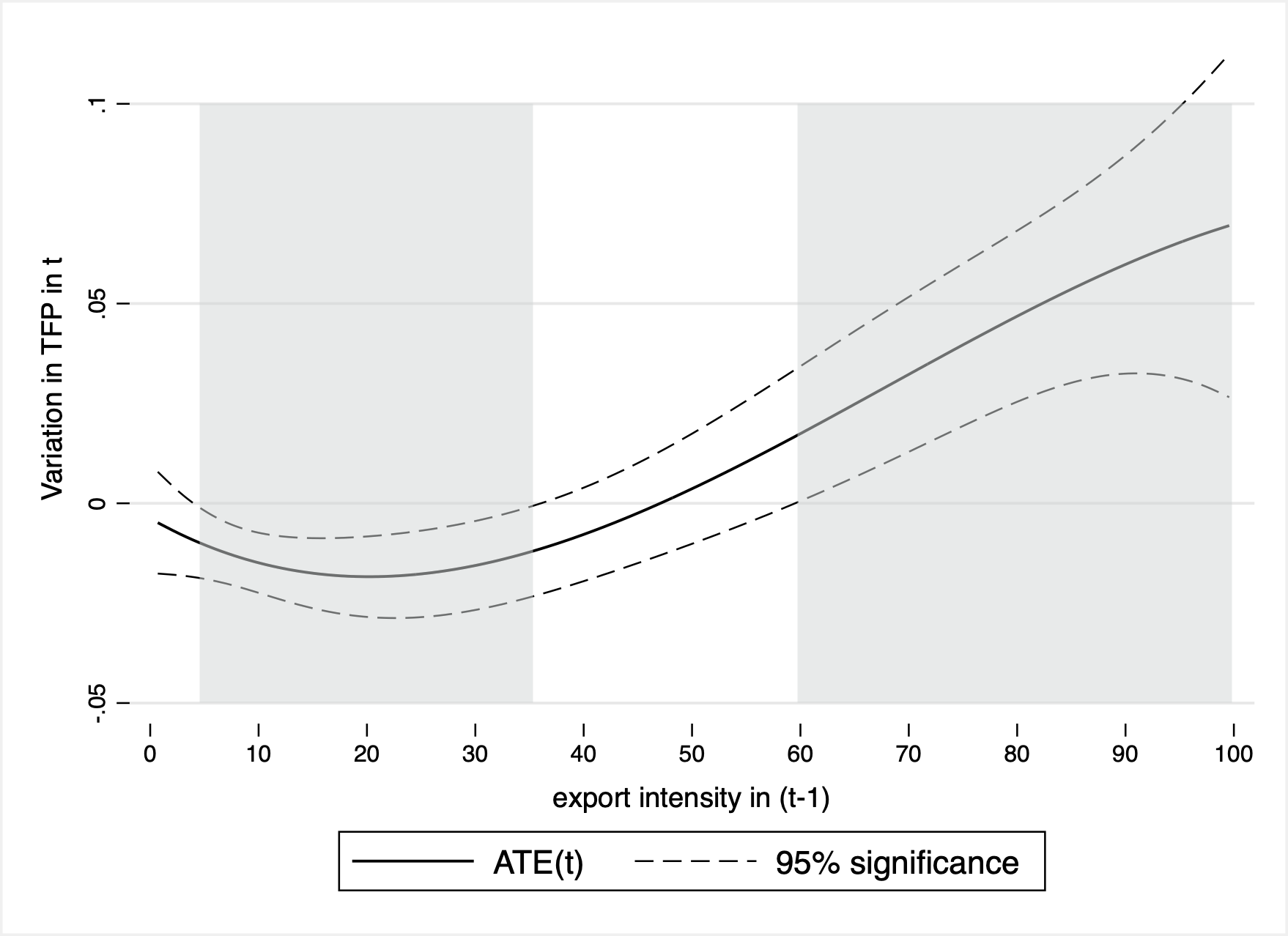}
    \begin{tablenotes}
    \singlespacing \footnotesize
    \item \textit{Note:} The figure reports the dose-response function obtained after estimating Eq. \ref{eq: drf}. The grey highlighted areas identify intervals of export intensity where the dose-response function is statistically different from zero using a 5\% significance level.
\end{tablenotes}
\end{figure}

In the following paragraphs, we present the results after estimating Eq. (\ref{eq: regression}). Our primary interest lies in studying the effect of export intensity on firm-level productivity. Central to our analysis is the dose-response function illustrated in Figure \ref{fig:drf_TFP}, which is obtained by plugging the coefficients from column (3) of Table \ref{tab:drf tfp expint} into Eq. (\ref{eq: drf}) and plotting the resulting curve over the export intensity support.

Please note how the inclusion of covariates, firm- and time-level fixed effects, contributes to improving the goodness of fit. Looking at our baseline estimates visualized in Figure \ref{fig:drf_TFP}, we can divide in segments. 

The first is the segment of an export intensity lower than 5\%. In this case, we do not find a significant impact on TFP. The typical firm may just be a passive exporter because the engagement is minimal. Such low export levels are insufficient for firms to benefit from either economies of scale or productivity spillovers. 

On the other hand, our findings show that an export intensity between 5\% and 35\% has, on average, a slight albeit statistically significant negative impact on TFP, which reaches its peak of $0.01\%$ at an export intensity of 20\%. We believe this is the segment where we find firms that become serious about exporting. They need to sustain sunk costs to organize the activities associated with a stable entry into foreign markets. Please note, however, that even if we find a relatively small reduction in TFP, it does not mean that the firm does not benefit from exporting. The firm still collects profits from a wider set of consumers at home and abroad. Yet, admittedly, between 5\% and 35\% of export intensity, we can find a great variety of situations. We are going to explore better what happens in single intervals in the next paragraphs, when we introduce additional firm-level outcomes for dose-response investigations.

Finally, we have the segment where export intensity is higher than 60\%, where a typical firm begins to reap productivity gains from exporting. Our intuition is that high levels of exporting induce economies of scale because the operational scale dramatically expands and knowledge spillovers emerge after intense interactions with foreign buyers. Our estimates indicate that, on average, firms experience a TFP rise ranging from 0.1\% to 0.6\% \textit{per year}, from the lower to the upper end of the range. Although apparently modest, the positive effects can accumulate over time.

\subsection{Other firm-level financial accounts}
\begin{figure}[htpb]
\centering
    \caption{The impact of export intensity on other firm-level outcomes}
    \label{fig:drf_channels}
    \subfloat[\label{subfig:a}Dose-response function of export intensity on Sales]{
    \includegraphics[width=0.45\textwidth]{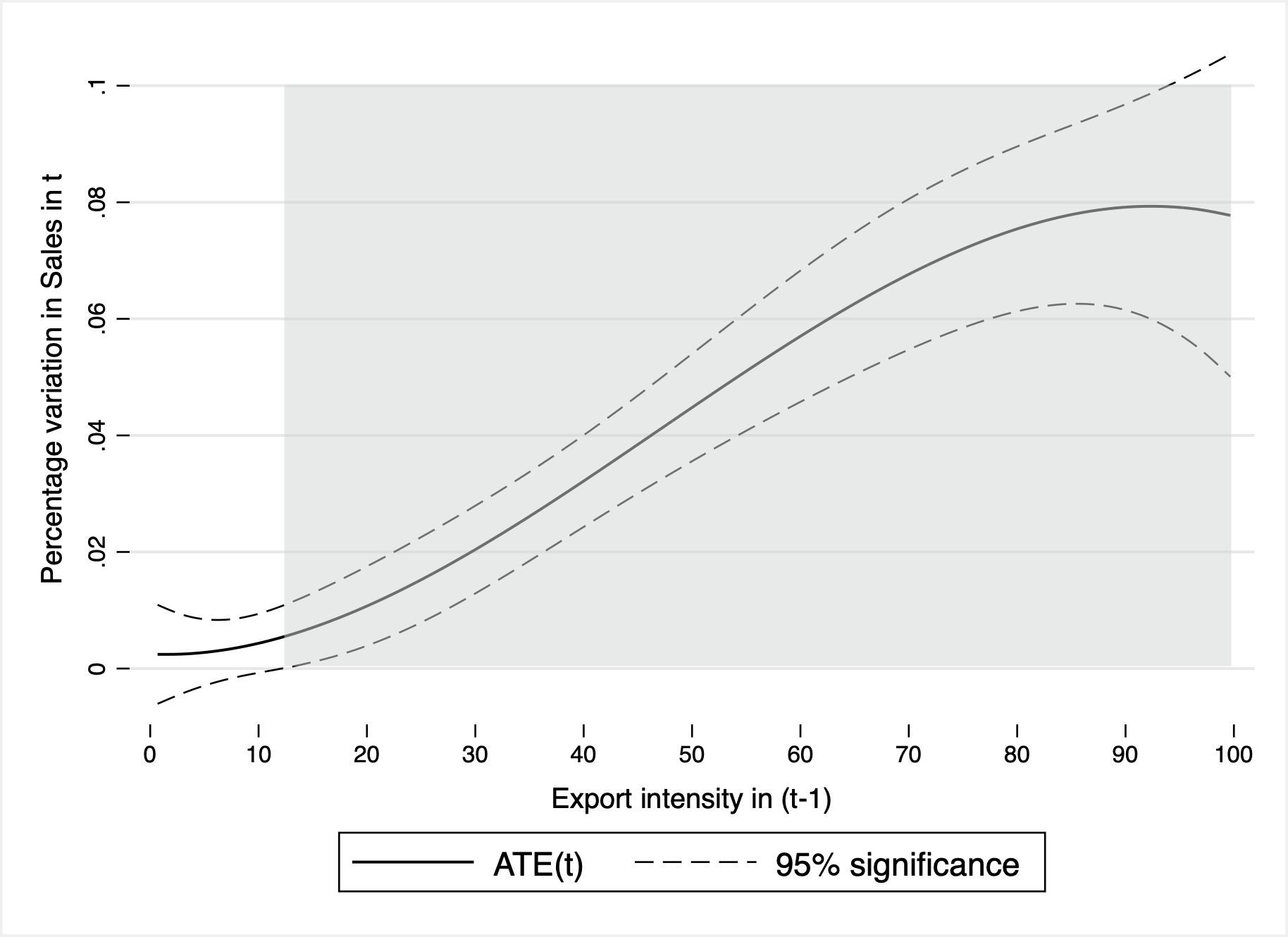}

  }%
  \hfill
  \subfloat[\label{subfig:b}Dose-response function of export intensity on Total Costs]{
    \includegraphics[width=0.45\textwidth]{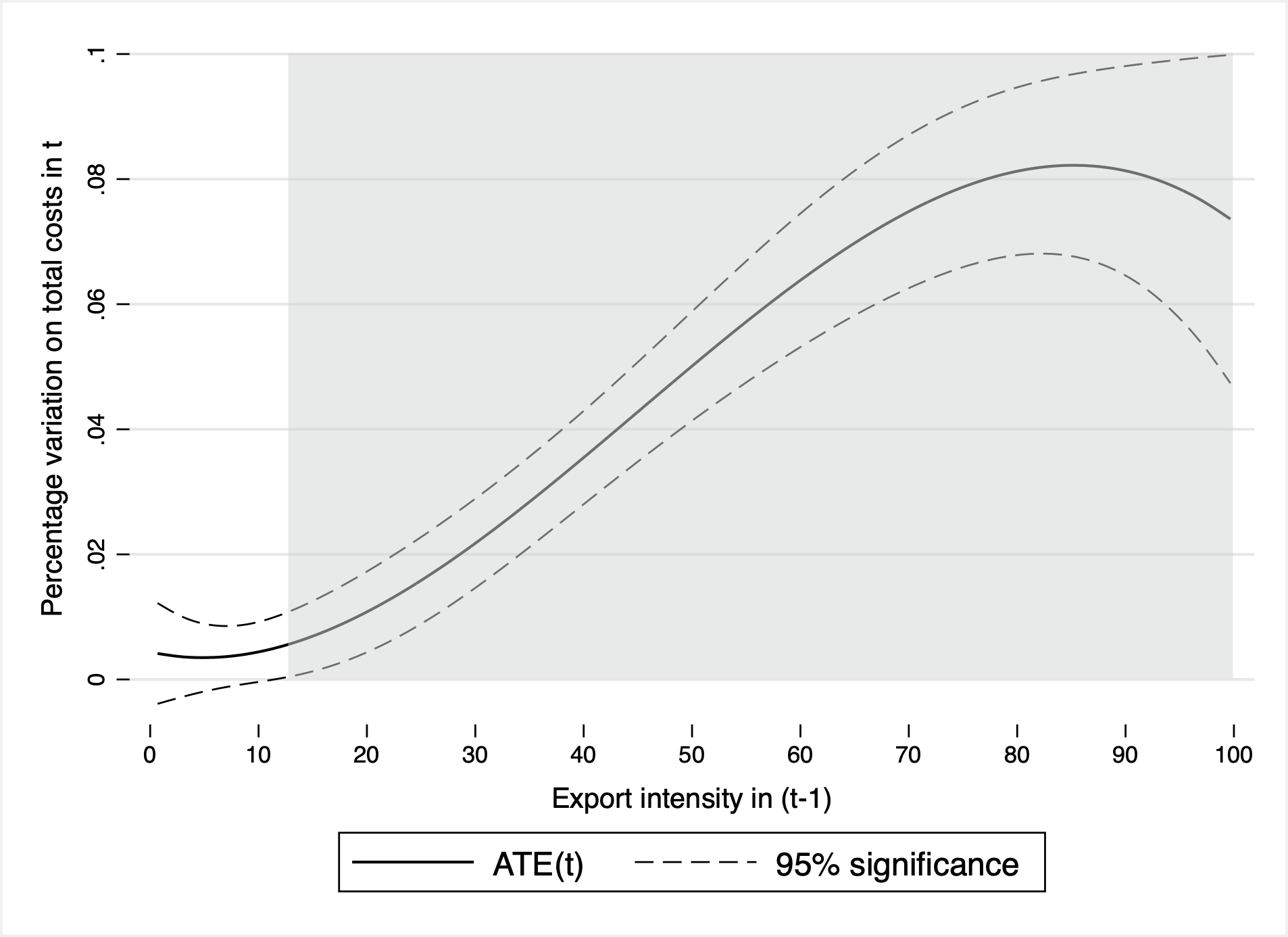}
 }
 \hfill
  \subfloat[\label{subfig:c}Dose-response function of export intensity on Capital intensity]{
    \includegraphics[width=0.45\textwidth]{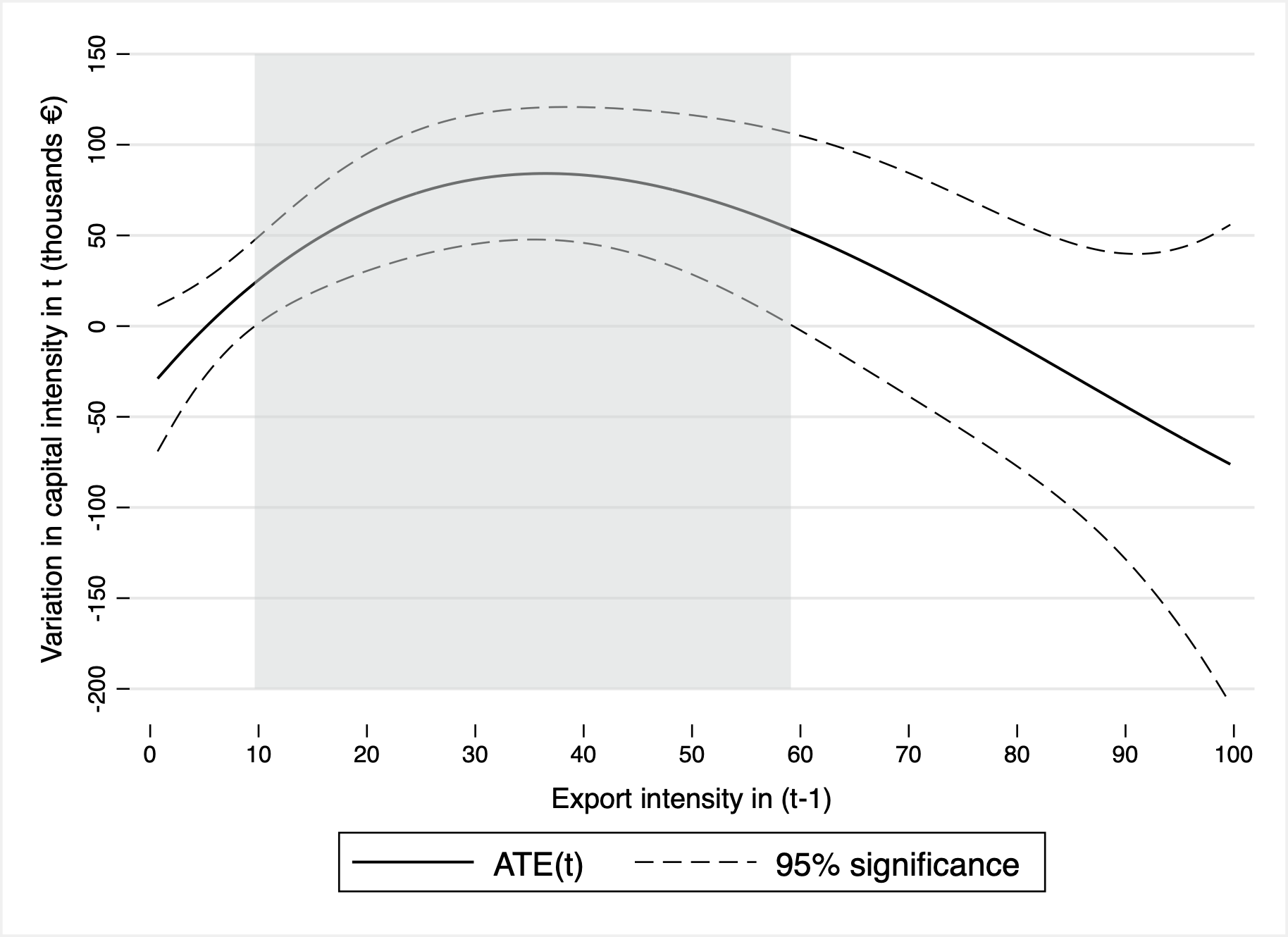}
 }
 \hfill
  \subfloat[\label{subfig:d}Dose-response function of export intensity on the Probability of filing a patent]{
    \includegraphics[width=0.45\textwidth]{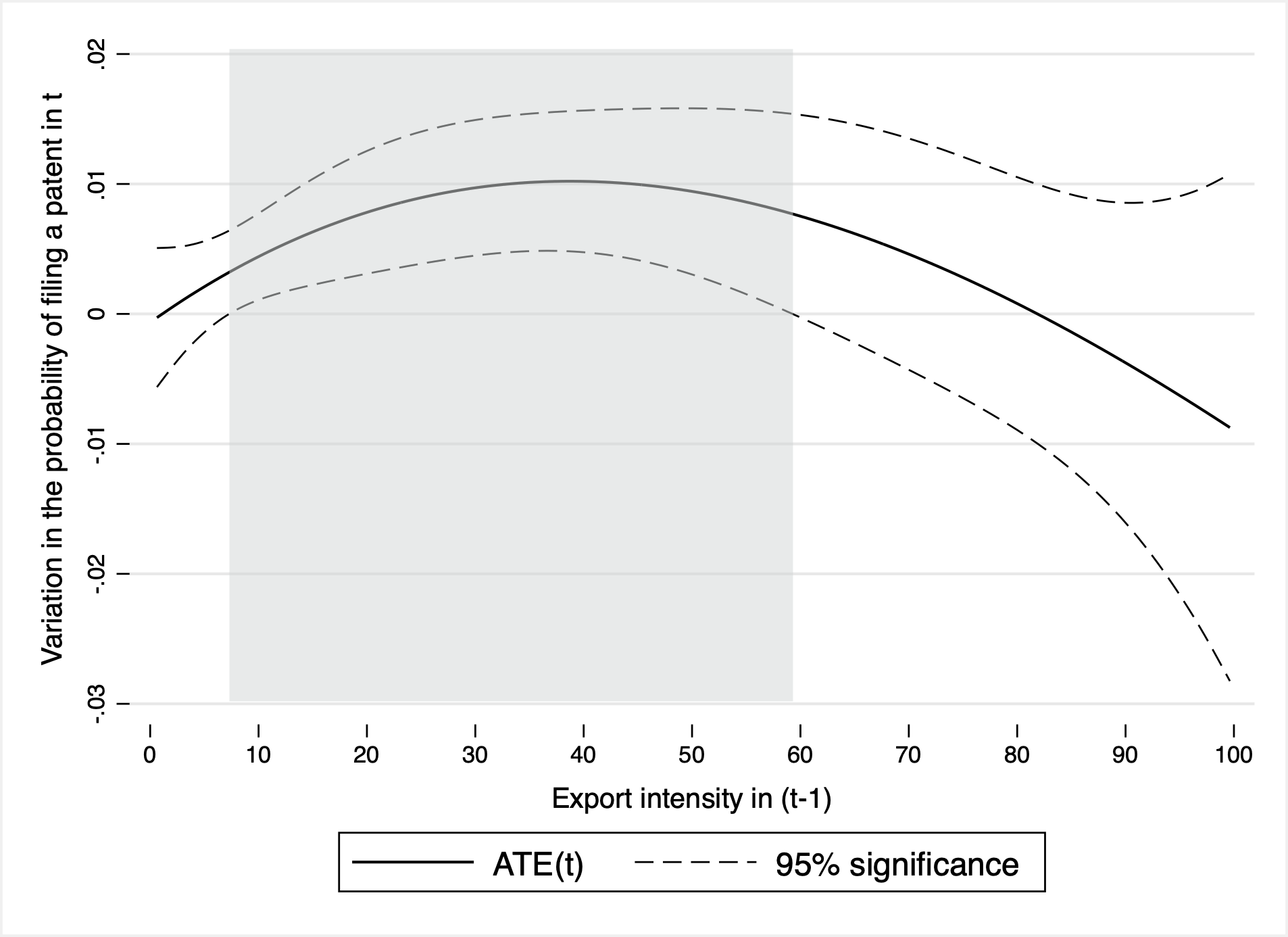}
 }\\
\begin{tablenotes}
    \singlespacing \footnotesize
    \item Note: The figures report the dose-response functions obtained after estimating Eq. \ref{eq: drf}. Figures (a), (b), (c), and (d)  show, respectively, the relationships between export intensity and (log of) sales, (log of) variable costs,  capital intensity, and the probability of filing a patent. The grey highlighted areas identify intervals of export intensity where the dose-response functions are statistically significant at the 95\% level.
\end{tablenotes}
\end{figure}

Previous segments in the relationship between TFP and export intensity can be explained by the prevalence of either of two different mechanisms. On the one hand, competition in international markets brings companies to catch up on the technological frontier and, thus, invest in new capital assets. On the other hand, a broader consumer base at home and abroad implies the emergence of economies of scale. For our purpose, we explore separately the impact of export intensity on firms' sales, variable costs, and capital intensity. In the next Section \ref{sec: technology}, we specifically address the impact of export intensity on the propensity to publish a patent. In Figure \ref{fig:drf_channels}, we visualize the shapes of the dose-response functions of all the alternative outcomes' equations, whose parameters are estimated in Table \ref{tab:reg_res}.

Notably, the dose-response functions for sales and variable costs in Figure \ref{fig:drf_channels} have an inverted U-shape that mirrors the one observed for TFP in \ref{fig:drf_TFP}. Our findings indicate that exporting significantly impacts costs and sales only when export intensity exceeds 10\%. Below this threshold, firms do not experience substantial changes in the operational scale. 

For firms with more than 10\% of activities destined for foreign markets, we observe a marked increase in operational volumes - they increase both sales and costs. Interestingly, however, sales increase more than costs, possibly pointing to economies of scale, whose presence is more evident when we record export intensities at 75\%, because costs peak while sales still increase. Eventually, confidence intervals blur with export intensity approaching the maximum of support.

The situation is different when we focus on capital intensity. The dose-response function reveals that exporting has a significant positive impact when intensities are in a 10\%–60\% range, and capital intensity peaks at about 32\%. The intuition is that the exporter needs to invest in new tangible and intangible fixed assets when the presence on foreign markets is stable. 

At this point, we can reconcile what we observe with the evidence from the last paragraphs on TFP in Figure \ref{fig:drf_TFP}. We argue that a critical mass is needed in either case to record a significant impact of the exporting activity. At lower levels of exporting activity, the company starts to benefit from economies of scale but also needs to invest in productive capacity. To keep up with the technological frontier is costly, and it often requires an upgrade of obsolete tangible assets. We argue that the combined evidence of rising operational capacity (sales and costs) and investment in fixed assets explains why we observe a negative albeit small productivity loss in an intermediate range of export intensity. It is only when the company operates abroad at a larger scale that positive albeit small TFP gains come as a consequence of exporting. In this case, we argue, economies of scale become evident and the capital adjustment unveils its impact on firms' performance.

\subsection{Export intensity and the probability to file a patent}\label{sec: technology}
Now we examine the impact of export intensity on the propensity to file a patent\footnote{There is an important strand of literature that studies the causal mechanisms linking patenting activity to firm performance. See for example \citet{Hedge_Luo_2018}, \citet{Farre-Mensa_et_al_2020}, and \citet{Exadaktylos_et_al_2024}. It is beyond the scope of this paper to unravel issues of reverse causality and endogenous innovation capabilities that are usually associated with patenting activities. Our scope is to understand whether segments of the dose-response curve generated by export intensity on productivity may find an association in alternative firm-level outcomes that potentially explain the local (on the segment) impacts.}. We estimate Eq. \ref{eq: drf}, where the outcome is the binary variable that takes a value of one if the firm registers a patent, and zero otherwise. In this case, the dose-response curve in Figure \ref{subfig:d} confirms that a minimum level of international exposure to exports is necessary to influence firms’ success in filing a patent. Specifically, no significant propensity increase is observed for export intensities below 8\%. Once this threshold is crossed, the average treatment effect becomes significantly positive, reaching its peak at around 35\% of export intensity. This suggests that moderate and sustained engagement with foreign markets yields the highest probability of patenting, likely corresponding to a phase in which firms consolidate their international presence and begin to reap strategic benefits from global competition, customers' feedback, and technological spillovers.

Beyond 60\% of export intensity, however, the positive effect on propensity diminishes and becomes statistically insignificant as firms approach a full exposure to exports. Please not that the shape of this curve closely mirrors that of capital intensity, fostering the intuition that innovation in exporting firms is deeply linked to the strategic accumulation of production and knowledge assets. The observed pattern reflects a key mechanism: when firms have a stable presence in international markets, they are more likely to invest in tangible and intangible fixed assets (such as new machinery, technologies, human capital, and organizational know-how) that support their long-term competitiveness abroad. We believe exporting and innovation appear to be jointly driven by a deeper capability-building process at the firm level.

\subsection{Common support and balance}
\label{subsec: common_support}

\begin{figure}[ht!]
\caption{Common support of the treated and the control observations}
    \centering
    \begin{subfigure}[b]{0.49\textwidth}
    \centering
    \includegraphics[width=\textwidth]{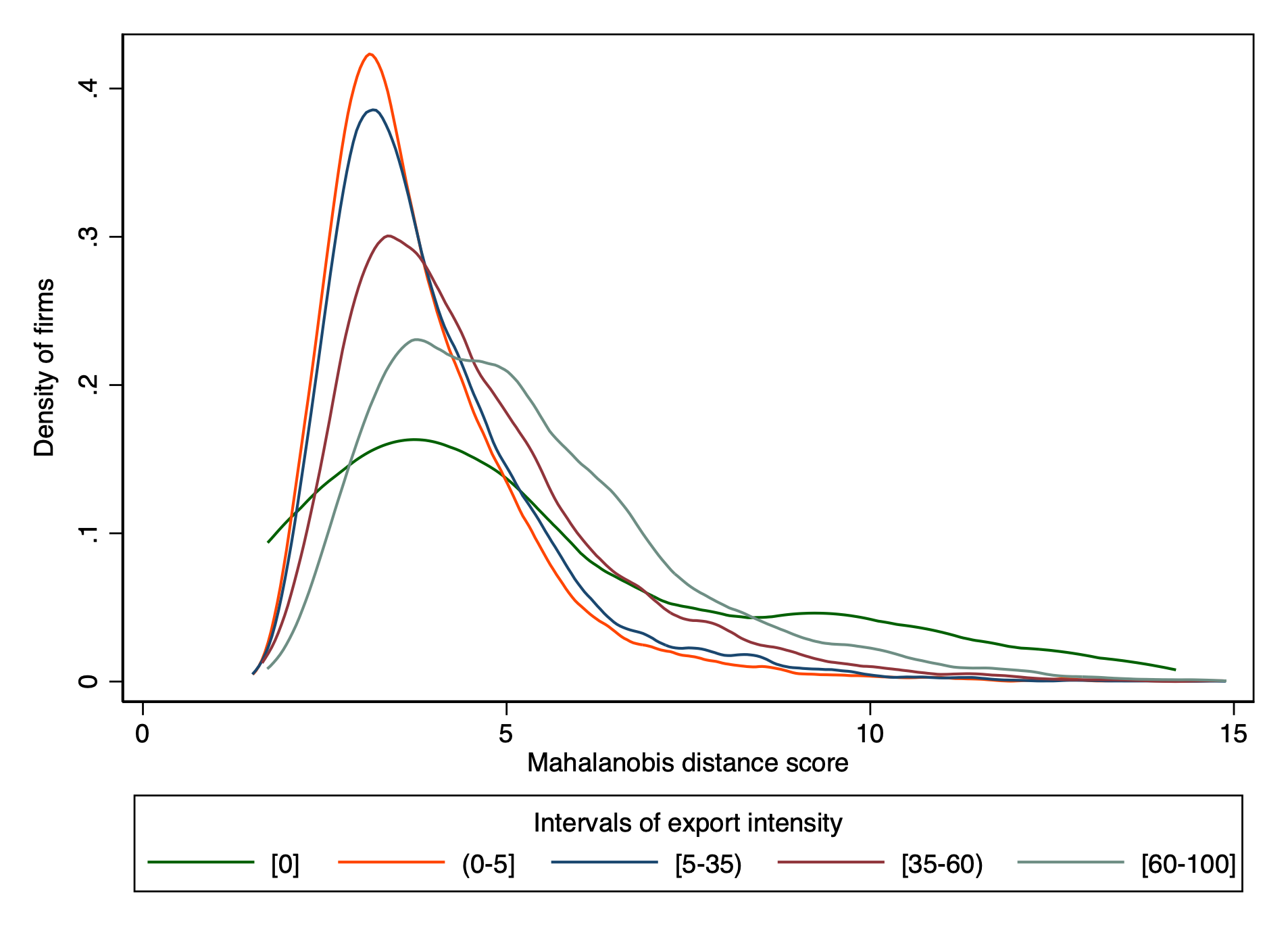}
    \caption{Distribution of Mahalanobis distances across  export intensity classes \label{fig:maha}}
    \end{subfigure}
    \begin{subfigure}[b]{0.49\textwidth}
    \centering
    \includegraphics[width=\textwidth]{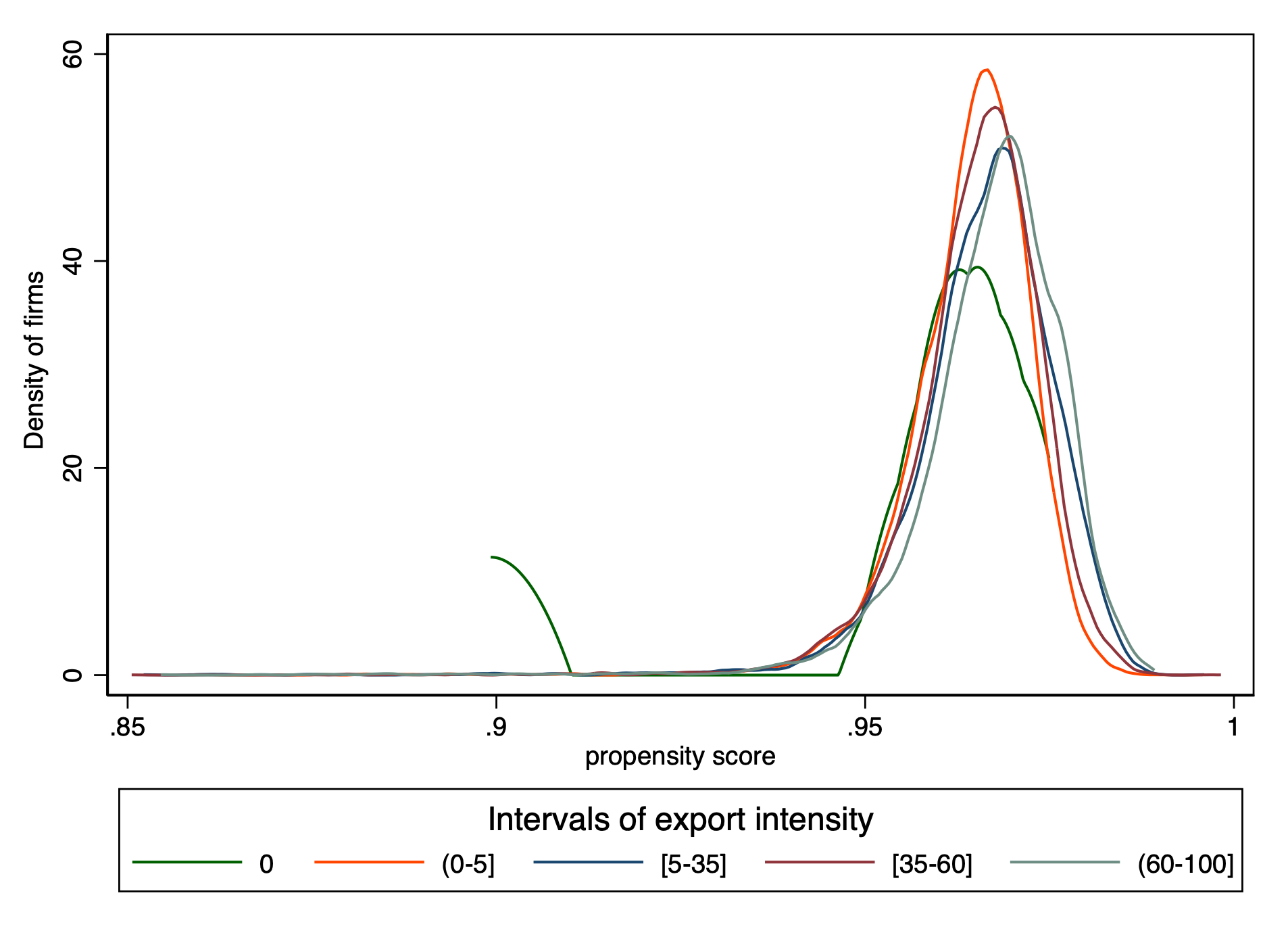}
    \caption{Distribution of propensity scores across export intensity classes\label{fig:pca}}
    \end{subfigure}
    \begin{tablenotes}
\singlespacing
\footnotesize
    \item Note: Figure (a) reports the distribution of Mahalanobis distances of firms in different export intensity classes, while Figure (b) reports the distribution of propensity scores of firms in different export intensity classes.
\end{tablenotes}
\end{figure}

In this section, we evaluate the quality of our counterfactuals. As far as we know, no standard procedure exists to check that randomization worked in a dose-response exercise like ours. We propose here to investigate whether the observations in the treated and the control groups have common distribution support and whether they balance in observable characteristics that can be endogenous to changes in export intensity. For our purpose, we start by computing propensity scores:

\begin{equation*}
    p(x_{it}) = P(D_{it} = 1 | X_{it-1} = x_{it-1})
\end{equation*}

where $D$ is the treatment indicator for a firm at time $t$, which takes the value one if the firm exports and zero otherwise, $X_{it-1}$ is the vector of observed covariates in the period before. We then examine how firms with varying export intensities are distributed across the propensity score distributions in Figure \ref{fig:pca}. We observe that distributions overlap for sample firms in the treated and control groups. For robustness, a similar exercise has been done by adopting Mahalanobis distances \footnote{In our context, a Mahalanobis distance ($M$) is a measure of dissimilarity between exporters and non-exporters. Let us consider a point $\mathbf{x}$ to a distribution characterized by a mean vector  $\boldsymbol{\mu}$ and a covariance matrix $\mathbf{\Sigma}$, then $M$ can be expressed as $M = \sqrt{(\mathbf{x} - \boldsymbol{\mu})' \mathbf{\Sigma}^{-1} (\mathbf{x} - \boldsymbol{\mu})}$, where $ \mathbf{x}$ is the vector of observations, $\boldsymbol{\mu}$ is the mean vector of the dataset, $\mathbf{\Sigma}$ is the covariance matrix of the dataset} distances instead of propensity scores. Results are reported in Appendix Figure \ref{fig:maha}. Finally, we compile tables of balancing properties stratified by quintiles (Appendix Tables \ref{tab:quintile1}–\ref{tab:quintile5}), therefore assuring that there is no fundamental difference between the treated and the control group when we consider firm-level characteristics that are possibly endogenous to the relationship between TFP and export intensity. Overall, the results demonstrate that most balancing conditions are met, with minor deviations observed in the first quintile.

\begin{figure}[H]
    \centering
    \caption{Dose-response function of export intensity on Total factor Productivity after a propensity score matching}
    \includegraphics[width=.7\linewidth]{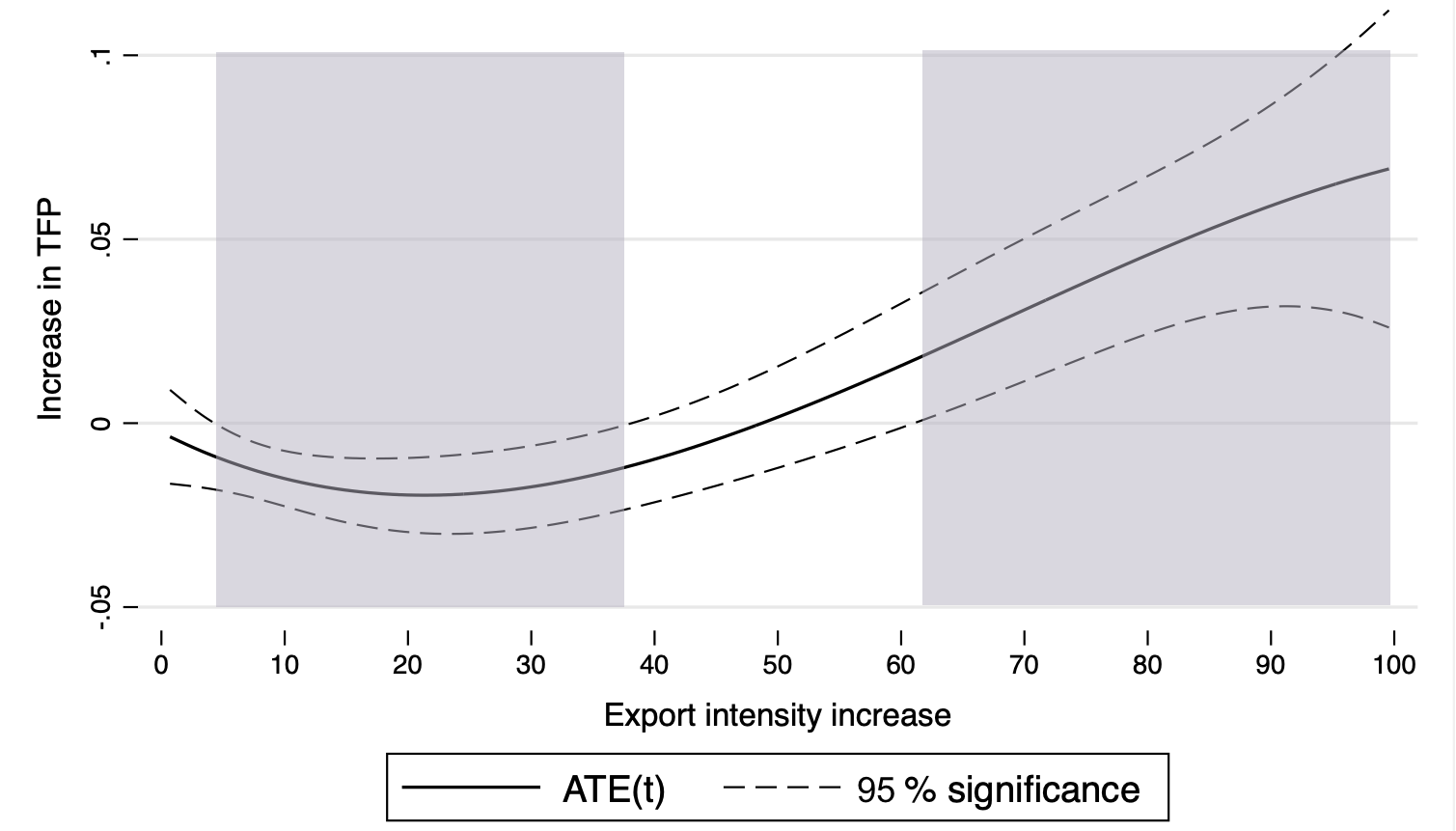}
    
    \label{fig:drfic_matched}
    \begin{tablenotes}
       \item  \footnotesize \singlespacing \textit{Note:} The figure shows the dose-response function obtained when using a control group derived from a nearest-neighbour propensity score matching. 
    \end{tablenotes}
\end{figure}

Finally, we re-estimate the dose-response function for the relationship between TFP and export intensity using the matched sample introduced above. Results are presented in Figure \ref{fig:drfic_matched}. Consistent with expectations, we confirm that the shape after matching is similar to that obtained from our baseline model.

\section{Robustness and sensitivity}
\label{sec: robustness}

In the next paragraphs, we address concerns about the robustness and sensitivity of our results. 

\begin{figure}[ht!]
    \caption{Dose-Response Functions - Robustness and sensitivity}
    \label{fig:drf_robust}
    \subfloat[All exporters\label{subfig:all_exporters}]{
    \includegraphics[width=0.48\textwidth]{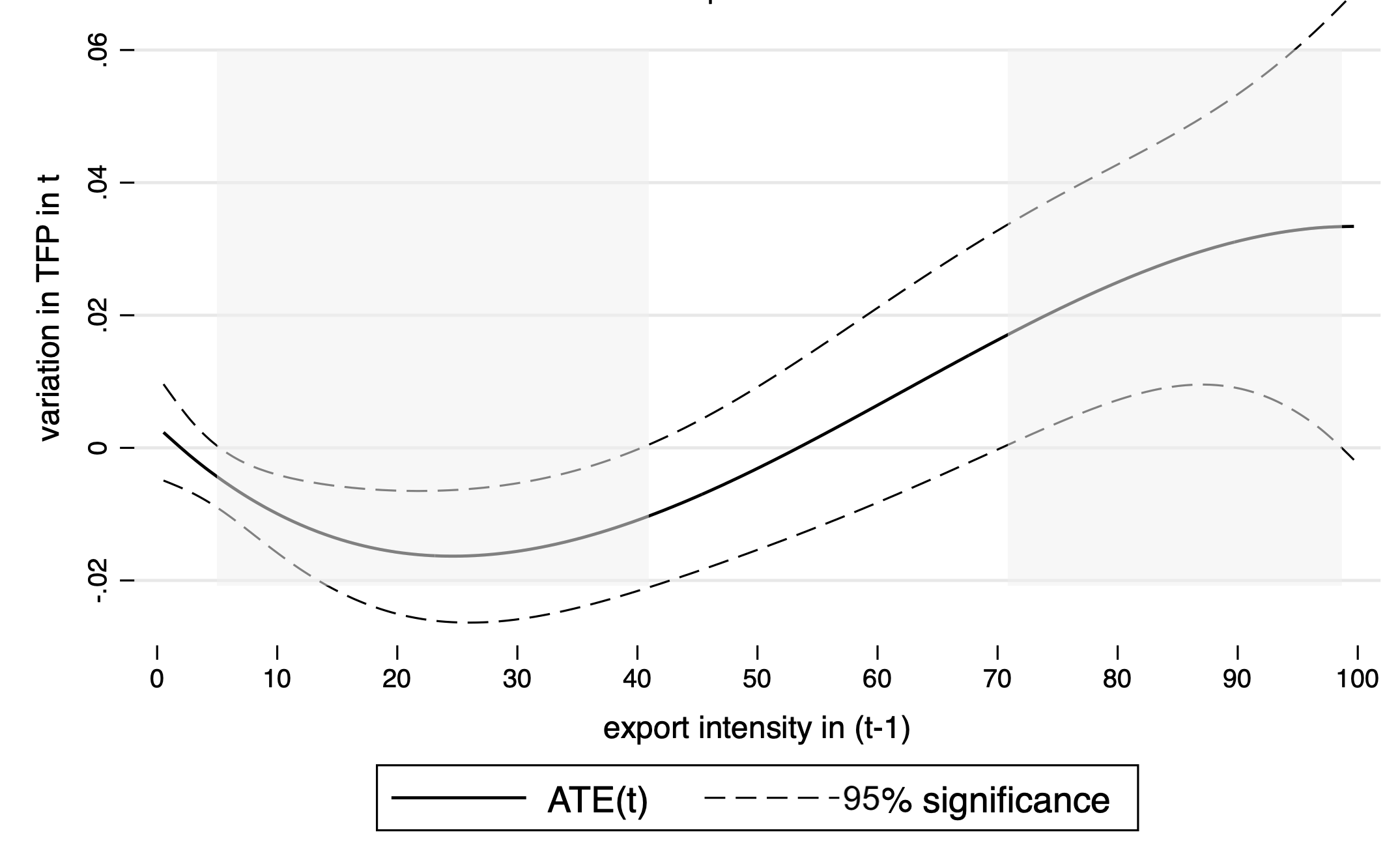}
}%
  \hfill
 \subfloat[Temporary exporters only\label{subfigtemp_expb}]{
 \includegraphics[width=0.48\textwidth]{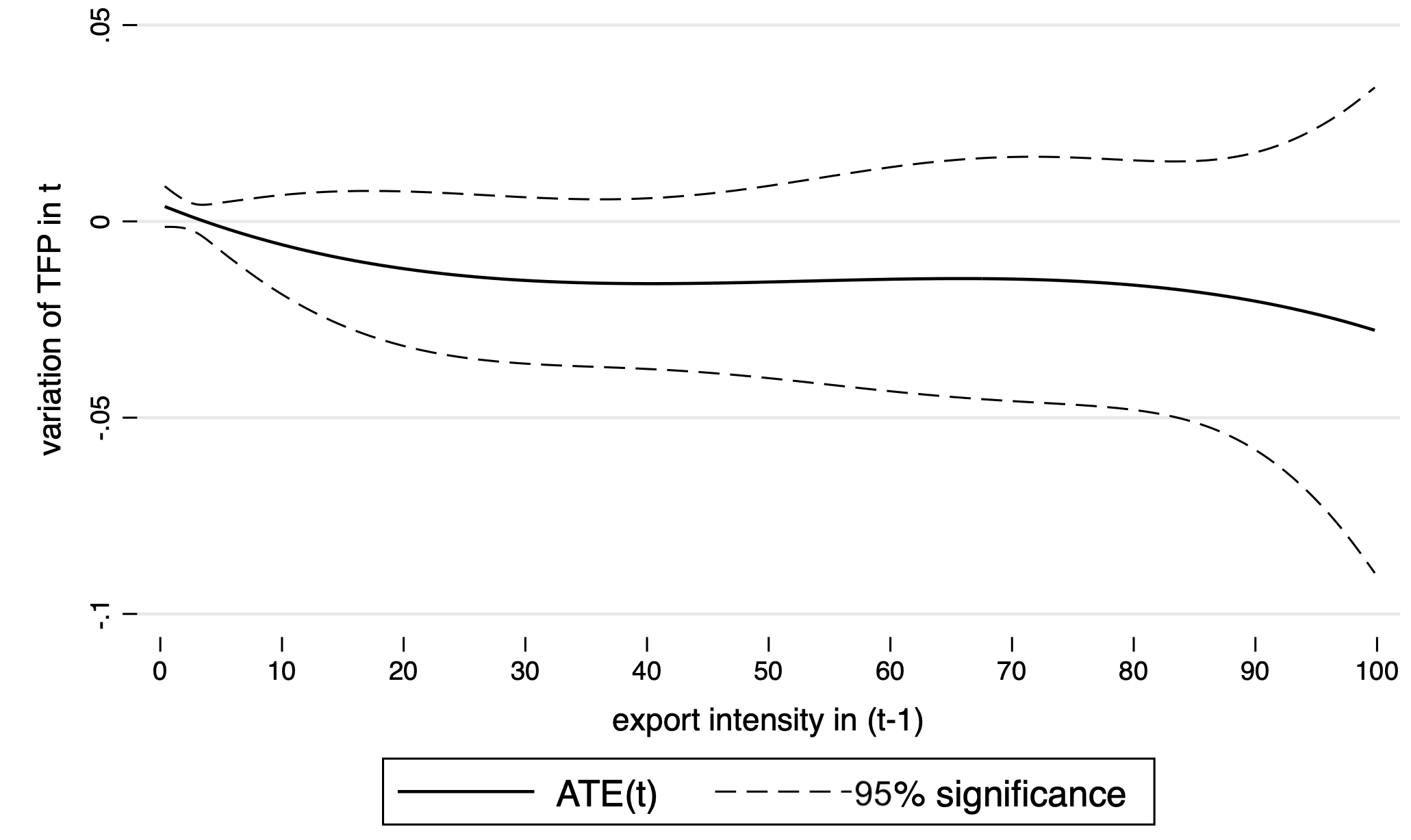}
  }\\
  \subfloat[Treatment in t-3\label{subfig:L3_treat}]{
    \includegraphics[width=0.48\textwidth]{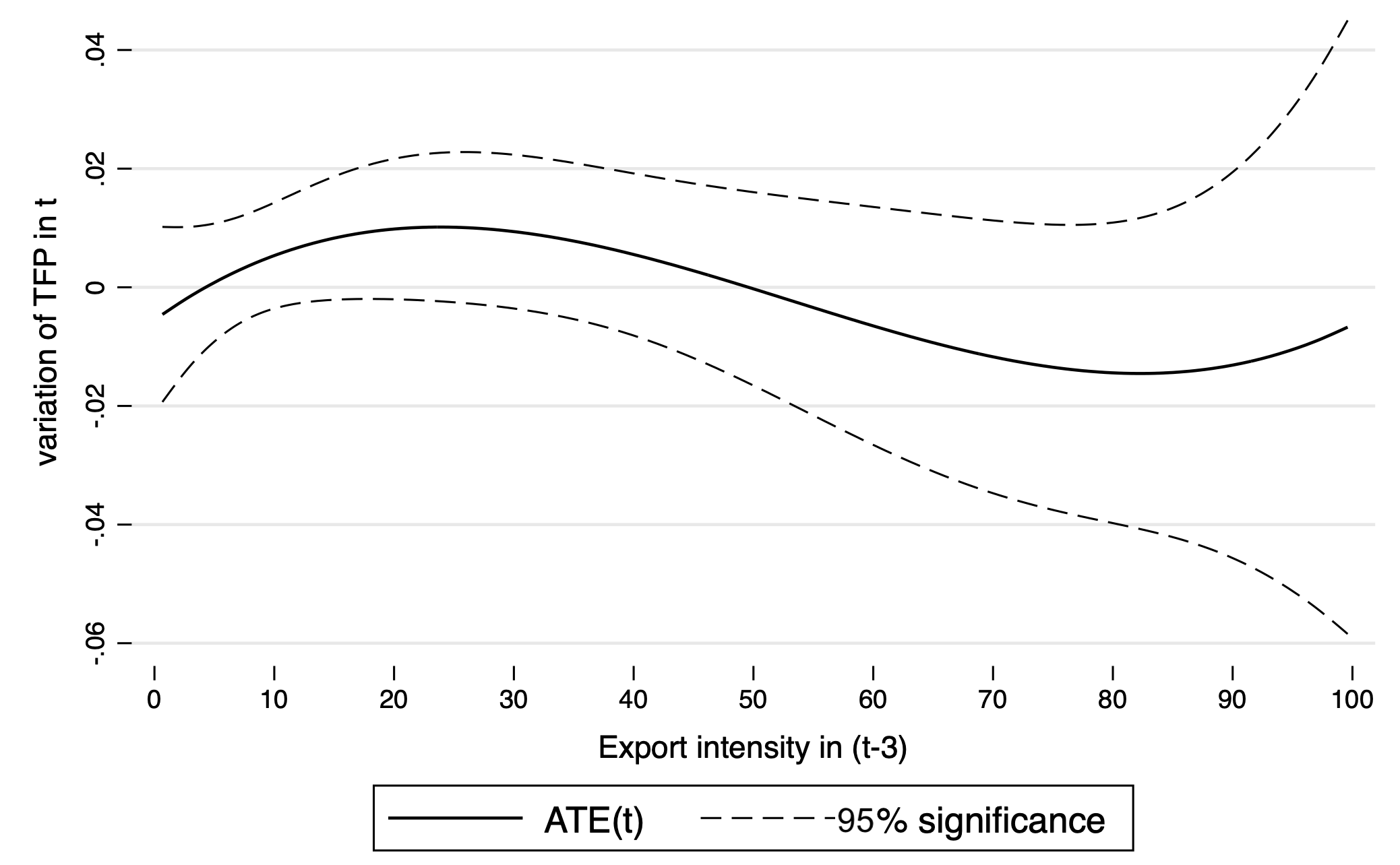}
  }%
  \hfill
  \subfloat[Including interaction terms\label{subfig:hetero_treat}]{
    \includegraphics[width=0.48\textwidth]{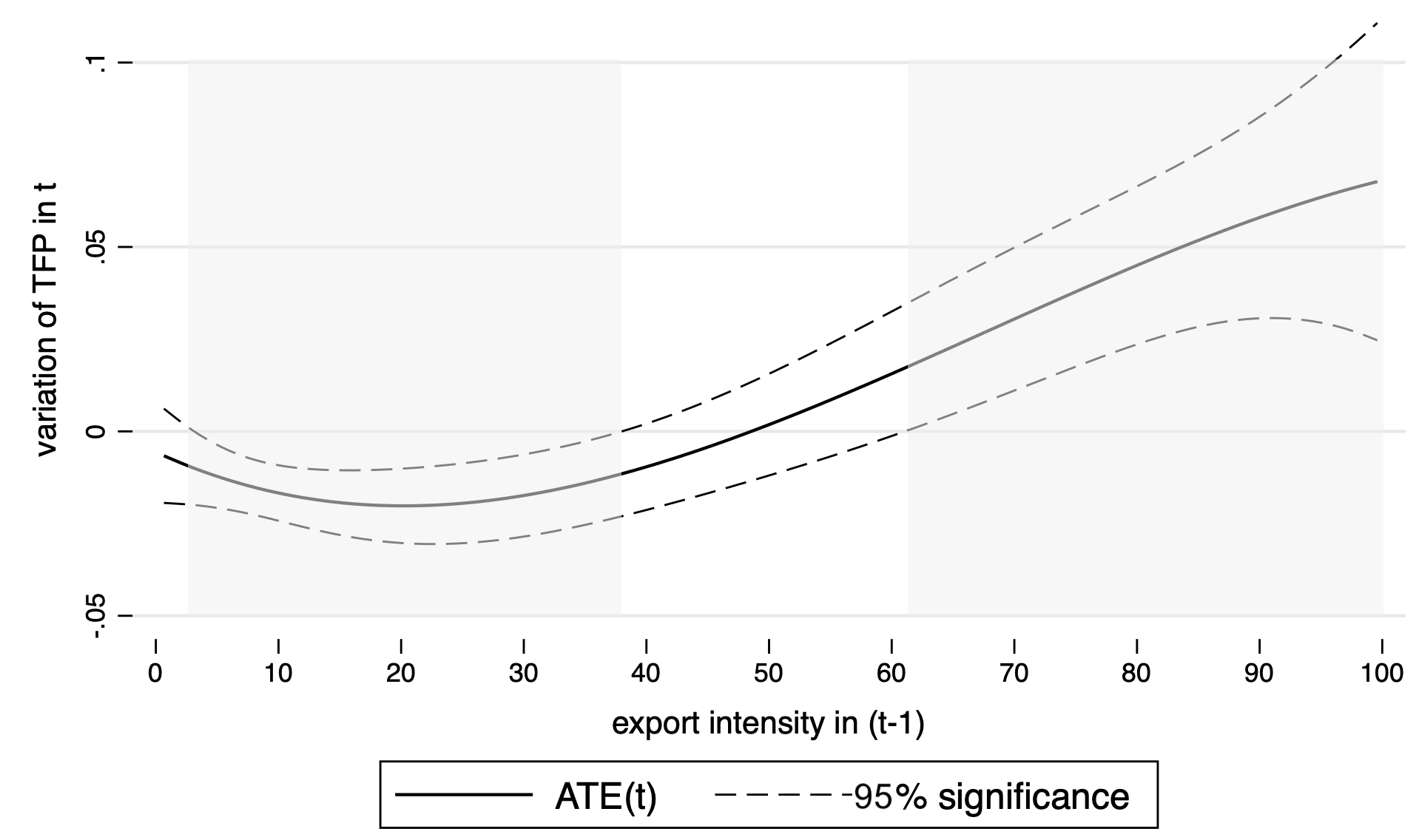}
}
\begin{tablenotes}
    \singlespacing \footnotesize
    \item Note: The figures report the dose-response functions for different robustness and sensitivity checks. Figure (a) reports the results after including temporary exporters in our sample, while Figure (b) reports the ones after considering only temporary exporters. In Figure (c) we introduce a three-year lag in the treatment of the export intensity. Figure (d) reports the dose-response function when we test for the incompleteness of the specification, including interaction terms. The grey highlighted areas identify intervals of export intensity where the dose-response function is statistically different from zero using a 95\% confidence interval.
\end{tablenotes}
\end{figure}

The first concern is that sample composition effects may drive our results, as we selected only non-temporary exporters from the beginning. A critique would be that we cherry-picked our sample. In Figure \ref{subfig:all_exporters}, we show that the shape of the dose-response function is robust to including temporary exporters in our analysis. On the contrary, and in line with our expectations, when we consider only temporary exporters, the effect of export intensity on a firm's productivity completely disappears. See Figure \ref{subfigtemp_expb}. This is in line with our expectations. 

A second concern relates to the duration of the effect of export intensity on a firm's productivity. One would argue that productivity benefits take some time to realize. Therefore, we ran a few specifications considering lags higher than one in the firm's exporting activity. Figure \ref{subfig:L3_treat} displays the dose-response function for exporting activity in $(t-3)$, where we can see no significant impact. The latter evidence suggests that learning-by-exporting mechanisms (if any) occur in the shorter term.

A further concern is that our set of controls might be incomplete, and the baseline equation might be underspecified. We added interaction terms between our controls and the varying treatment. The dose-response function remains unaffected. See Figure \ref{subfig:hetero_treat}.

An additional concern relates to the choice of the functional form of $h(d)$. Our baseline is a polynomial of third-degree, but we experimented with alternative polynomials, ranging from a linear specification (degree 1) to a fifth-degree. Across these specifications, the dose-response functions systematically reveal a negative effect of export intensity on TFP within the 5\%-35\% range, and a positive effect for export intensities above 60\%. The only exception is the linear model, which, however, identifies an export intensity of approximately 30\% as a critical threshold. Notably, the main cutoffs identified by our baseline exercise are robust to different choices of the functional form. See Appendix Table \ref{app_tab:Poly_TFP} and Appendix Figure \ref{app_fig:Poly_TFP} for further details.

\begin{figure}[htpb!]
\centering
    \caption{Export intensity and TFP: firms in border regions}
    \label{fig: multidest}
     \subfloat[\label{subfig:border} All regions at the borders]{\includegraphics[width=0.45\textwidth]{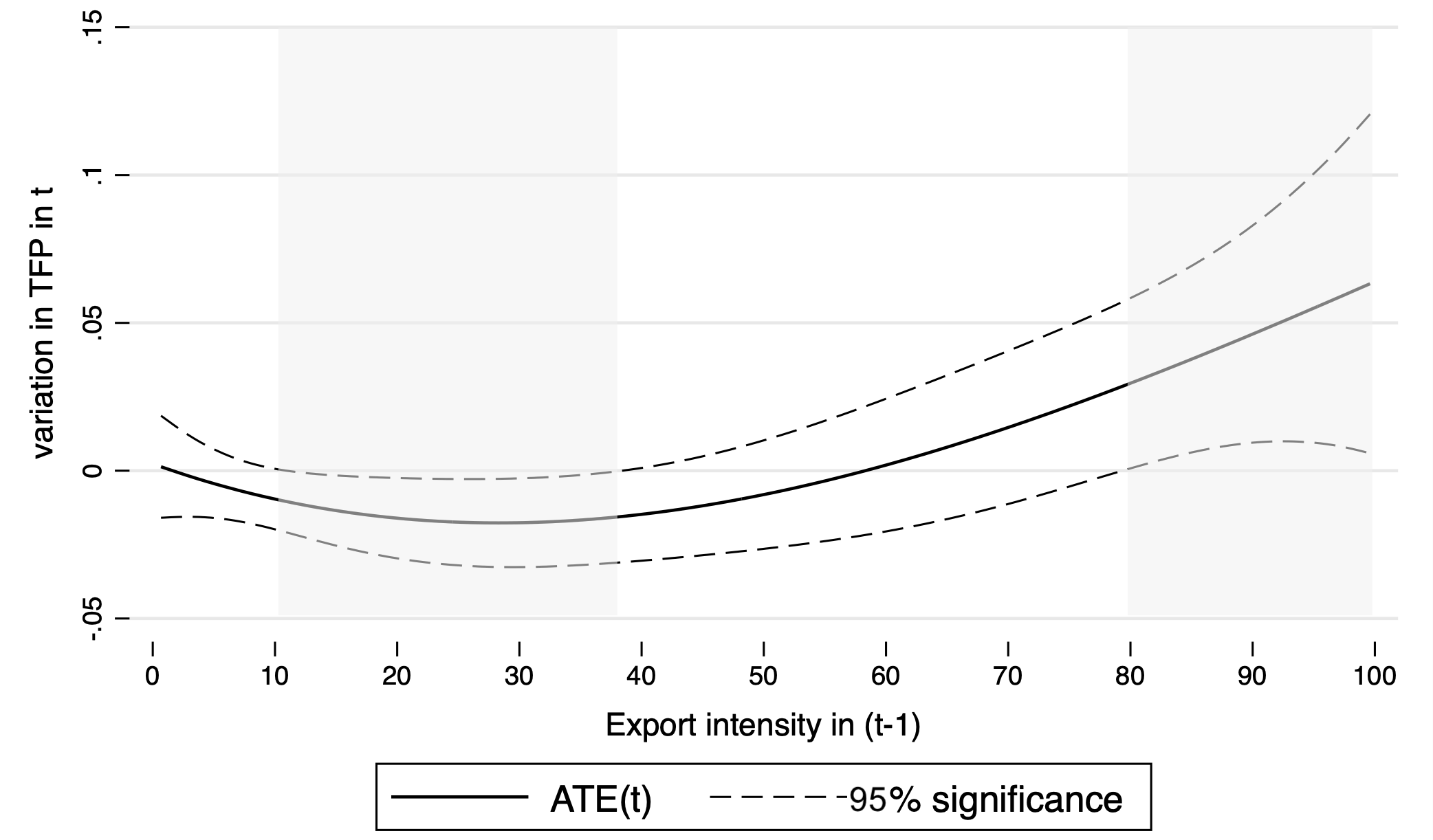}}  
        \subfloat[\label{subfig:ys_FE}Including industry-year fixed effects]{\includegraphics[width=0.45\textwidth]{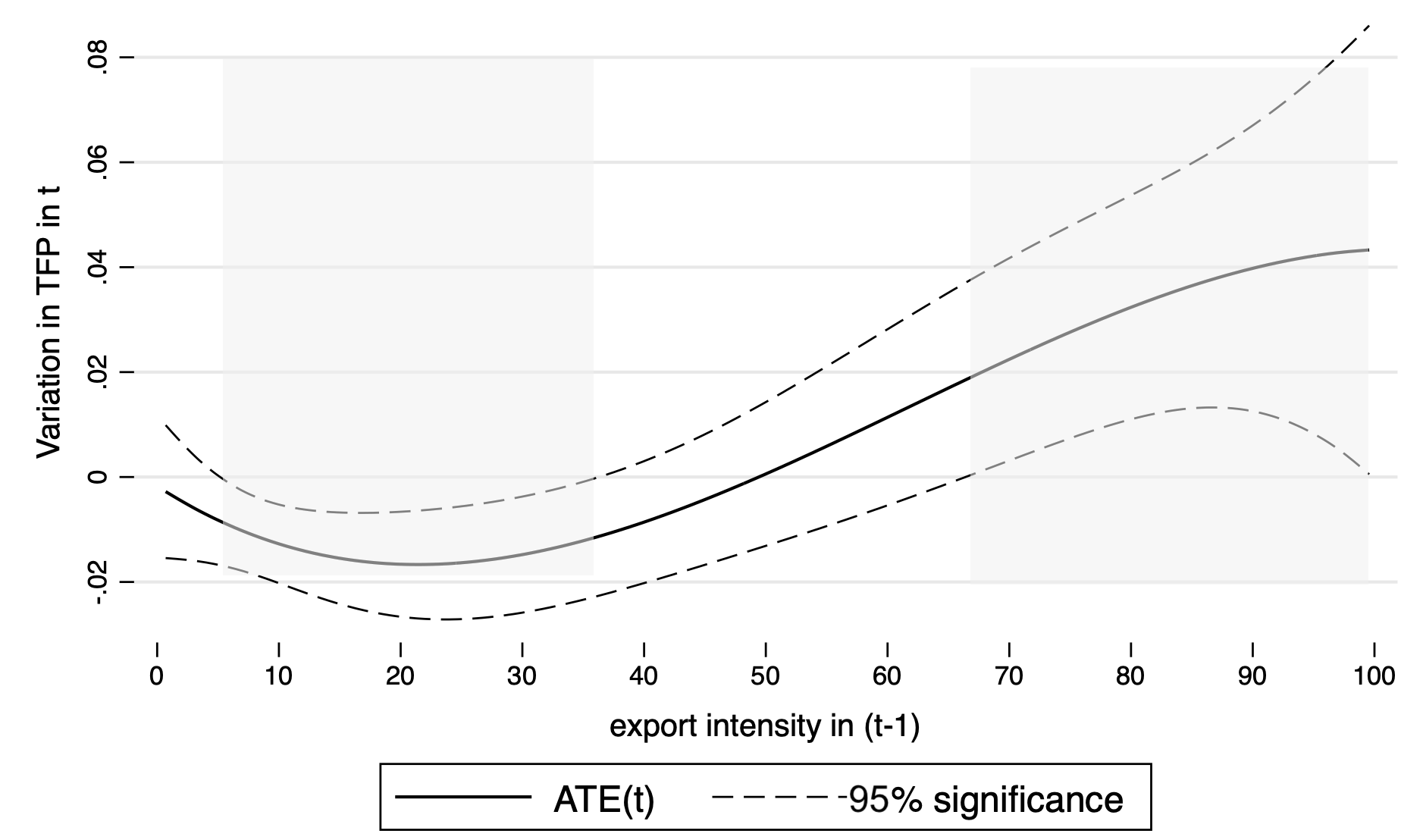}}
    \begin{tablenotes}
        \singlespacing \footnotesize
        \item \textit{Note:} The Figure reports in panel (a) the estimated dose-response function we obtain when restricting our sample to all French regions at the border with other countries; panel (b) reports the estimated dose-response function we obtain when including industry-year fixed effects.
    \end{tablenotes}
\end{figure}

Another concern relates to the lack of information about export destinations. See also further discussion in the next Section \ref{sec: limitations}. In this case, we propose a sensitivity check focusing on exporters located in border regions. Our intuition is that exporters at the border are different from exporters in internal regions. Exporters in border regions are more likely to export to geographically proximate countries. In this case, increases in export intensity are more likely to stem from intensive margin expansions and less from destination scope if compared with firms located in internal regions. Eventually, the dose-response function we obtain in Figure \ref{subfig:border} shows that the relationship between export intensity and productivity remains consistent with our baseline findings, suggesting that the role of destination diversification in shaping our results is moderate.

A final concern relates to the lack of information about firm-level demand, as it can play a crucial role \citep{DeLoecker_2013}\footnote{See also \citet{Aw.2011,almunia2021venting}.}. Although our data structure does not allow for identifying the specific markets in which the firms operate, we perform a robustness check by introducing industry-year fixed effects to account for sector-specific shocks in a given year. The resulting dose-response function in Figure \ref{subfig:ys_FE} shows that the relationship between export intensity and productivity remains consistent with our baseline findings.

\section{Limitations}\label{sec: limitations}
An important limitation of our data is the lack of information on export destinations and on firms' product portfolios. Destination diversification has often been related to risk reduction and learning opportunities \citep{esposito2022demand} because it provides firms with access to knowledge about market competitors and consumer preferences \citep{eaton2004dissecting,de2007exports}. In our identification strategy, firm-level fixed effects absorb the distinction between multi-destination and single-destination exporters, thus controlling for the fact that productivity gains or losses are contingent upon the firm’s ability to access a broader pool of experiences. What remains unverified is the composition effect of adding destinations or products to the firm's portfolio. Our results do not tell us whether the impact can be decomposed into an intensive or extensive margin.

Another limitation stems from the lack of information on export prices. In our identification strategy, a deflation of firm-level outcomes with producer price indices and the inclusion of year-specific fixed effects moderates the impact of industry-varying and time-varying inflation. Yet, at the firm level, we cannot observe composition effects due to changing prices and quantities of products. 

Please note how, despite the previously described composition effects, our findings in magnitudes and statistical significance cannot be biased by unobservable characteristics unless they simultaneously correlate with the dependent variable and the treatment, i.e., with firm-level productivity and export intensity. Omitted factors affecting only one of these dimensions would not bias our findings.

\section{Conclusions}\label{sec: conclusion}

This paper explores the causal relationship between export intensity and firms' financial accounts using a dose-response framework. By considering export intensity as a continuous treatment, we examine how varying degrees of presence in foreign markets influence firm performance, while controlling for self-selection into exporting and firm-level heterogeneity. Our first interest is on productivity, to contribute to the literature that investigates the causal mechanism between exporting and productivity. Then, we investigate the economic mechanisms by exploring the relationship between export intensity on the one hand and sales, costs, capital intensity, and registered patents on the other hand.

Our findings reveal a non-linear relationship between export intensity and productivity. At low levels of export intensity (below 5\%) the impact on productivity is negligible, likely reflecting minimal or passive export activities. Between 5\% and 35\% of export intensity, firms experience a small but statistically significant decline in productivity. It is at this stage that we observe movements in other firm-level outcomes: sales increase faster than variable costs, on the one hand, capital intensity and the propensity to file patents increase, on the other hand. It is only when export intensity exceeds 60\% that firms begin to reap small but statistically significant productivity gains, ranging from 0.1\% to 0.6\% per year. 

Our findings are robust across alternative sample selections, model specifications, and identification strategies. Nonetheless, some limitations remain, as we lack detailed information on export destinations and product portfolios, which could unravel possible composition effects from prices and quantities.

Overall, our results suggest that only firms reaching a substantial degree of internationalization are able to reap learning-by-doing effects. This has important policy implications: efforts to support firm internationalisation should not only focus on helping firms initiate export activities, thus working on the extensive margin. Policymakers should can work on the intensive margin by helping exporting firms to sustain the sunk costs needed to scale up their presence in foreign markets and unlock the full benefits of learning-by-exporting.

\pagebreak

\onehalfspacing
\setlength\bibsep{0.5pt}
\bibliographystyle{elsarticle-harv}
\bibliography{biblio}
\newpage

\section*{Appendix: Tables and graphs}
\setcounter{table}{0}
\renewcommand{\thetable}{A\arabic{table}}
\setcounter{figure}{0}
\renewcommand{\thefigure}{A\arabic{figure}}

\begin{small}
\begin{longtable}{p{.45\textwidth}p{.45\textwidth}}
  \caption{Variables}
  \label{app_tab: variables}\\
\hline \multicolumn{1}{l}{\textbf{Variable}} & \multicolumn{1}{l}{\textbf{Description}}  \\ \hline 
\endfirsthead

\multicolumn{2}{c}%
{{\bfseries \tablename\ \thetable{} -- continued from previous page}} \\
\hline \multicolumn{1}{l}{\textbf{Variable}} & \multicolumn{1}{l}{\textbf{Description}} \\ \hline 
\endhead

\hline \multicolumn{2}{r}{{Continued on next page}} \\ \hline
\endfoot

\hline \hline
\endlastfoot
Sales, Number of employees, Profit Margins, P/L after tax, Operating revenue turnover, Working capital, Long-term debt, Debtors, Tangible fixed assets, Intangible fixed assets, Financial Expenditure &  Original financial accounts expressed in euro.\\
Export intensity & Indicator computed as \textit{Export revenues/ Total revenues}\\
Total Costs& Total costs of production, computed as \textit{Real Cost of materials +Real Cost of employees}\\

Profitability & Measure of profitability expressing how much earnings are generated by the firm's assets. It is computed as  \textit{EBITDA/Total Assets}\\
NACE rev. 2 & A 2-digit industry affiliation following the European Classification\\
NUTS 2-digit & The region in which the company is located following the European classification.\\
TFP & It is the Total Factor Productivity of a firm computed as in \citet{ackerberg2015identification}.\\
Size-Age &  It is a synthetic indicator proposed by Hadlock and
Pierce (2010), computed as (-0.737$\cdot$ \textit{log}(\textit{total assets}) )+(0.043 $\cdot$ \textit{log}(\textit{total assets}))$^2$ -(0.040 $\cdot$ \textit{age}) to catch the non-linear relationship between financial constraints, size and age. \\
patents & It is a binary variable with value 1 if the firm possesses at least one patent at time \textit{t}\\
D(export in t-1) & It is a binary variable with value 1 if the firm reported positive export revenues in \textit{t-1}\\
Pavitt Class & Taxonomy, which describes a firm's patterns of technical change. The classification follows the methodology of \citet{bogliacino2016pavitt}, which is based on Nace Rev.2 classification.\\
Inward FDI & It is a binary variable with value 1 if the firm has foreign headquarters \\
Outward FDI & It is a binary variable with value 1 if the firm has subsidiaries abroad\\
N.patents & Total number of patents owned by the firm at time \textit{t}\\
Corporate Control & A binary variable equals one if a firm belongs to a corporate group.\\
Labour Productivity & It is a ratio between value added and number of employees for the average productivity of labor services.\\
Productive Capacity & It is an indicator of investment in productive capacity computed as \textit{Fixed Assets}$_t$/(\textit{Fixed Assets}$_{t-1}$+\textit{Depreciation}$_{t-1}$)\\
Capital Adequacy Ratio &  It is a ratio of Shareholders' Funds over Short and Long Term Debts. \\
Financial Sustainability & It is a ratio between Financial Expenses and Operating Revenues.\\
Capital Intensity & It is a ratio between fixed assets and number of employees for the choice of factors of production.\\
Petent's filing & It is a dummy that takes value 1 if the firm filed a patent in t, 0 otherwise\\
Firm Size & Size classification sourced from Orbis:\begin{itemize}
    \item \textit{Very Large}: they match at least one of the following conditions:
\begin{itemize}[noitemsep]
\item Op. revenue $\geq$100 million €
\item Total assets $\geq$ 200 million €
\item Employees $\geq$ 1,000
\item Listed
\end{itemize}
    \item \textit{Large:} they match at least one of the following conditions:
\begin{itemize}[noitemsep]
\item  Op. revenue $\geq$ 10 million € 
\item  Total assets $\geq$ 20 million €
\item  Employees $\geq$ 150
\item  Not very large
\end{itemize}
    \item \textit{Medium:} when they match at least one of the following conditions:
\begin{itemize}[noitemsep]
\item  Ope. revenue $\geq$ 1 million €
\item  Total assets $\geq$ 2 million €
\item  Employees $\geq$ 15
\item  Not very large or large
\end{itemize}
    \item \textit{Small:} Residual Class
\end{itemize}\\
\hline
\end{longtable}
\end{small}

\newpage
\begin{table}[]
    \centering
    \caption{Export Premium in Firm Productivity}
    \label{tab: export_premium_stats}
       \resizebox{.8\textwidth}{!}{
    \begin{tabular}{lccccccc}
    \toprule
   &Mean&St. dev.&p10&p25&p50&p75&p90\\
    \midrule
    Non exporters&9.900&0.800&9.136&9.471&9.804&10.240&10.765\\
Exporters&10.179&0.960&9.226&9.624&10.003&10.552&11.415\\
&&&&&&&\\
\hline
Total&10.099&0.926&9.192&9.571&9.942&10.451&11.22823\\
\bottomrule
    \end{tabular}}
\begin{tablenotes}
\footnotesize \singlespacing
    \item \textit{Note:} The table presents summary statistics on firms' Total Factor Productivity (TFP) for exporters and non-exporters following the production function approach by \citet{ackerberg2015identification}.
\end{tablenotes}
\end{table}

\begin{table}[htb!]
    \caption{Estimates of dose-response parameters for TFP vs. export intensity}
    \label{tab:drf tfp expint}
    \centering
   
    \begin{tabular}{lc@{\hskip 0.5in}cc}
    \toprule
     & TFP & TFP & TFP \\
     &(1)&(2)&(3)\\
    \midrule
    $D_1$& 3.31e-3* & -1.78e-3 & -1.53e-3 \\
    & (1.38e-3) & (9.54e-4) & (9.05e-4) \\
    $D_2$& 1.37e-4*** & 4.31e-5 & 4.46e-5 \\
    & (3.86e-5) & (2.58e-5) & (2.45e-5) \\
    $D_3$& -1.09e-6*** & -1.88e-7 & -2.19e-7 \\
    & (2.82e-7) & (1.87e-7) & (1.77e-7) \\
\midrule
Firm FE & No & Yes& Yes \\
Time FE & No & Yes& Yes\\
Controls & No & No & Yes\\
\midrule
N. obs. & 39,365& 39,365&39,365\\
R squared &0.038&0.001&0.101\\
RMSE &1.014&0.313&0.297\\
\bottomrule
    \end{tabular}
    \begin{tablenotes}
        \singlespacing
        \footnotesize
        \item Note: The table reports nested estimates of Equation \ref{eq: regression} using as dependent variable firm-level Total Factor Productivity (TFP) following \citet{ackerberg2015identification}. Controls include (log of) firm size measured by number of employees, financial constraints measured by the so-called size-age indicator proposed by \citet{hadlock2010new}, export status in the previous year, and the number of patents as a proxy of innovation ability. Standard errors are clustered at the firm level in parentheses (*** p$<$0.01, ** p$<$0.05, * p$<$0.1). 
    \end{tablenotes}
\end{table}

\begin{table}[htb!]
    \caption{The impact of export intensity on alternative outcomes}
    \label{tab:reg_res}
    \centering
    \resizebox{.8\textwidth}{!}{
    \begin{tabular}{lc@{\hskip 0.5in}ccc}
    \toprule
     &Sales&Costs&Capital intensity&Patent's filing\\
     &(1)&(2)&(3)&(4)\\
    \midrule
$D_1$ & -8.19e-05 & -3.79e-04 & 7.106* &5,92e-04  \\
& (6.04e-04) & (5.73e-04) & (2.866) &(4.09e-04)\\
$D_2$ & 2.88e-05 & 4.11e-05** & -1.24e-01 & -8.88e-06\\
& (1.63e-05) & (1.54e-05) & (7.74e-02) &(1.12e-05)\\
$D_3$ & -2.05e-07 & -3.04e-07** & 4.85e-04 &2.13e-08\\
& (1.17e-07) & (1.11e-07) & (5.58e-04) &(8.16e-08)\\
\midrule
Firm FE & Yes & Yes& Yes& Yes \\
Time FE & Yes & Yes& Yes& Yes\\
Controls & Yes & Yes & Yes& Yes\\
\midrule
N. obs &43,160  &43,162&43,925 &45,079 \\
R-squared & 0.172 &0.205&0.047&0.00107 \\
RMSE &0.214&0.203& 1,026.704&0.147\\
\bottomrule
    \end{tabular}}
    \begin{tablenotes}
        \singlespacing
        \footnotesize
        \item Note: The table reports the estimates of Eq. (\ref{eq: regression}) with dependent variables: (log of) sales, (log of) costs, and capital intensity expressed in thousand euro. In column (4), the outcome is a binary that takes value 1 if the firm has filed at least a patent in t, and  TFP is used as a control, together with the size-age measure and the log of number of employees.
    \end{tablenotes}
\end{table}

\begin{table}[H]
    \centering
        \caption{Export intensity and TFP: different polynomial specifications of $h(d)$}
    \label{app_tab:Poly_TFP}
    \resizebox{\textwidth}{!}{
    \begin{tabular}{lccccc}
    \toprule
    &TFP&TFP&TFP&TFP&TFP\\
&(1)&(2)&(3)&(4)&(5)\\
    \midrule
$D_1$         & 0.000659\sym{***}&-0.000591         & -0.00153\sym{*}  & -0.00104         & -0.00593\sym{***}\\
                &(0.000184)         &(0.000485)         &(0.000905)         &(0.00148)         &(0.00221)         \\
$D_2$          &                  &0.0000151\sym{***}&0.0000446\sym{*}  &0.0000170         & 0.000447\sym{***}\\
                &                  &(0.00000543)         &(0.0000245)         &(0.0000702)         &(0.000160)         \\
$D_3$      &                  &                  &-0.000000219         &0.000000268         &-0.0000128\sym{***}\\
                &                  &                  &(0.000000177)         &(0.00000118)         &(0.00000454)         \\

$D_4$      &                  &                  &                  &-2.63e-09         &0.000000158\sym{***}\\
                &                  &                  &                  &(6.30e-09)         &(5.43e-08)         \\
$D_5          $&                  &                  &                  &                  &-6.84e-10\sym{***}\\
                &                  &                  &                  &                  &(2.29e-10)   \\
\midrule
Polynomial degree h(t)&1&2&3&4&5\\
firm FE&YES&YES&YES&YES&YES\\
year FE&YES&YES&YES&YES&YES\\
Exporters&non-temporary&non-temporary&non-temporary&non-temporary&non-temporary\\
\midrule
(N)&39,365&39,365&39,365&39,365&39,365\\
\bottomrule

    \end{tabular}
    }
\begin{tablenotes}
    \singlespacing \footnotesize
    \item \textit{Note:} In this table we report the results of the model estimated exploring polynomial degrees from 1 (linear case) up to degree 5.
\end{tablenotes}
\end{table}

\begin{figure}[H]
    \caption{Dose-response functions - alternative polynomial specifications for $h(t)$}
    \label{app_fig:Poly_TFP}
    \begin{subfigure}[(a)]{0.49\textwidth}
     \includegraphics[width=\textwidth]{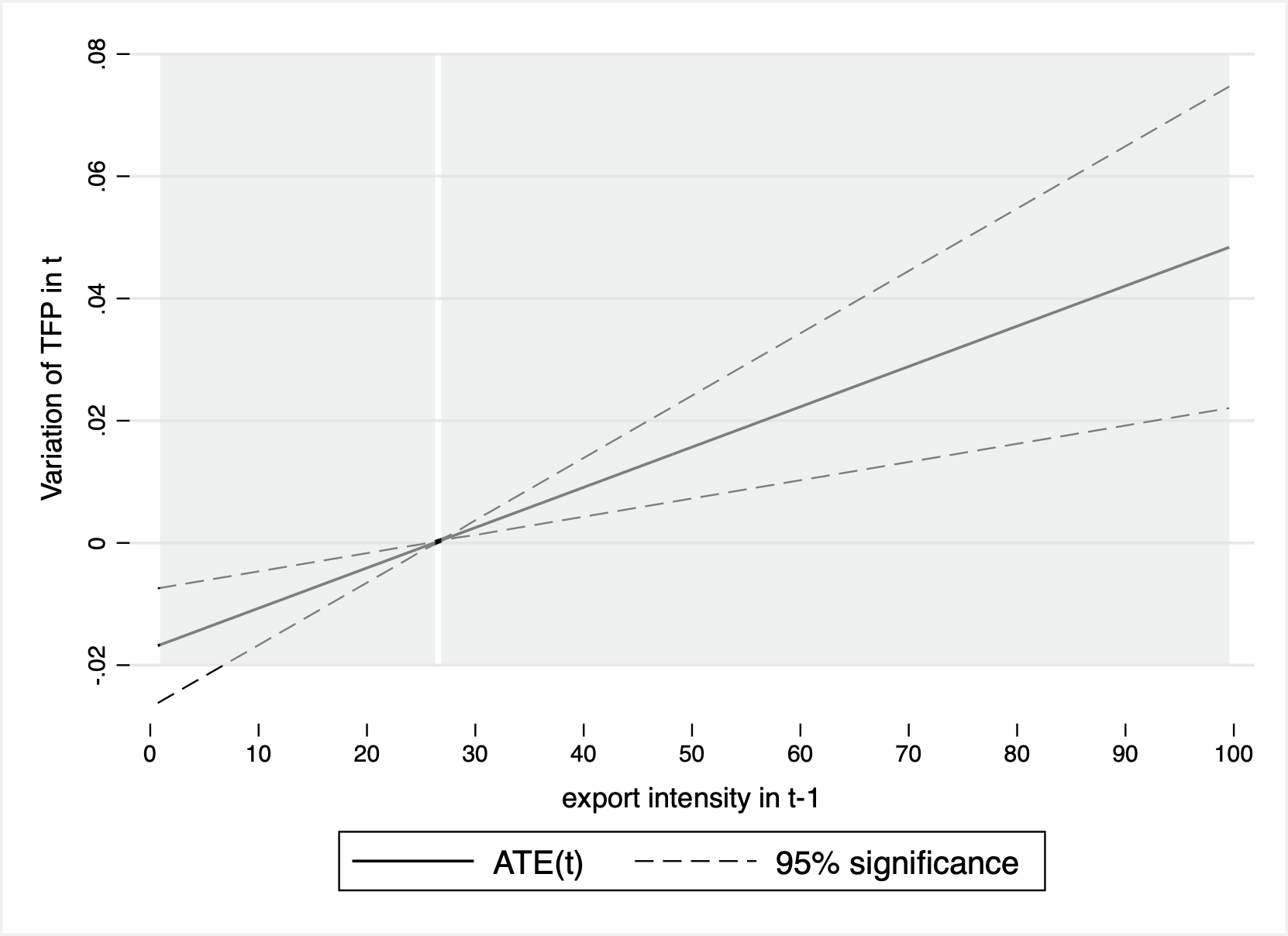}   
     \caption{Polynomial degree 1}
    \end{subfigure}
    \hfill
    \begin{subfigure}[(b)]{0.49\textwidth}
        \includegraphics[width=\textwidth]{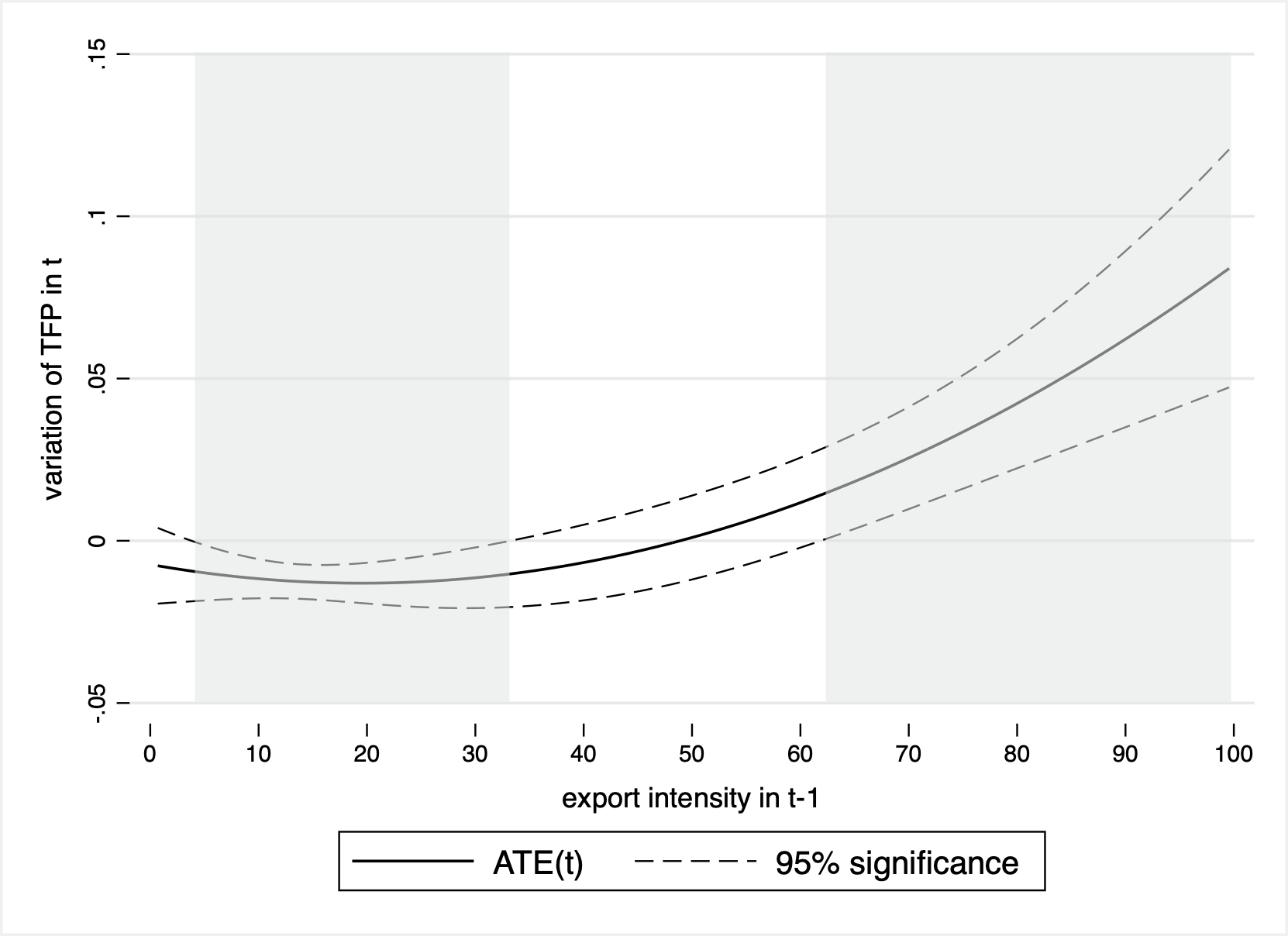}  
        \caption{Polynomial degree 2}
    \end{subfigure}\\
    \begin{subfigure}[(b)]{0.49\textwidth}
        \includegraphics[width=\textwidth]{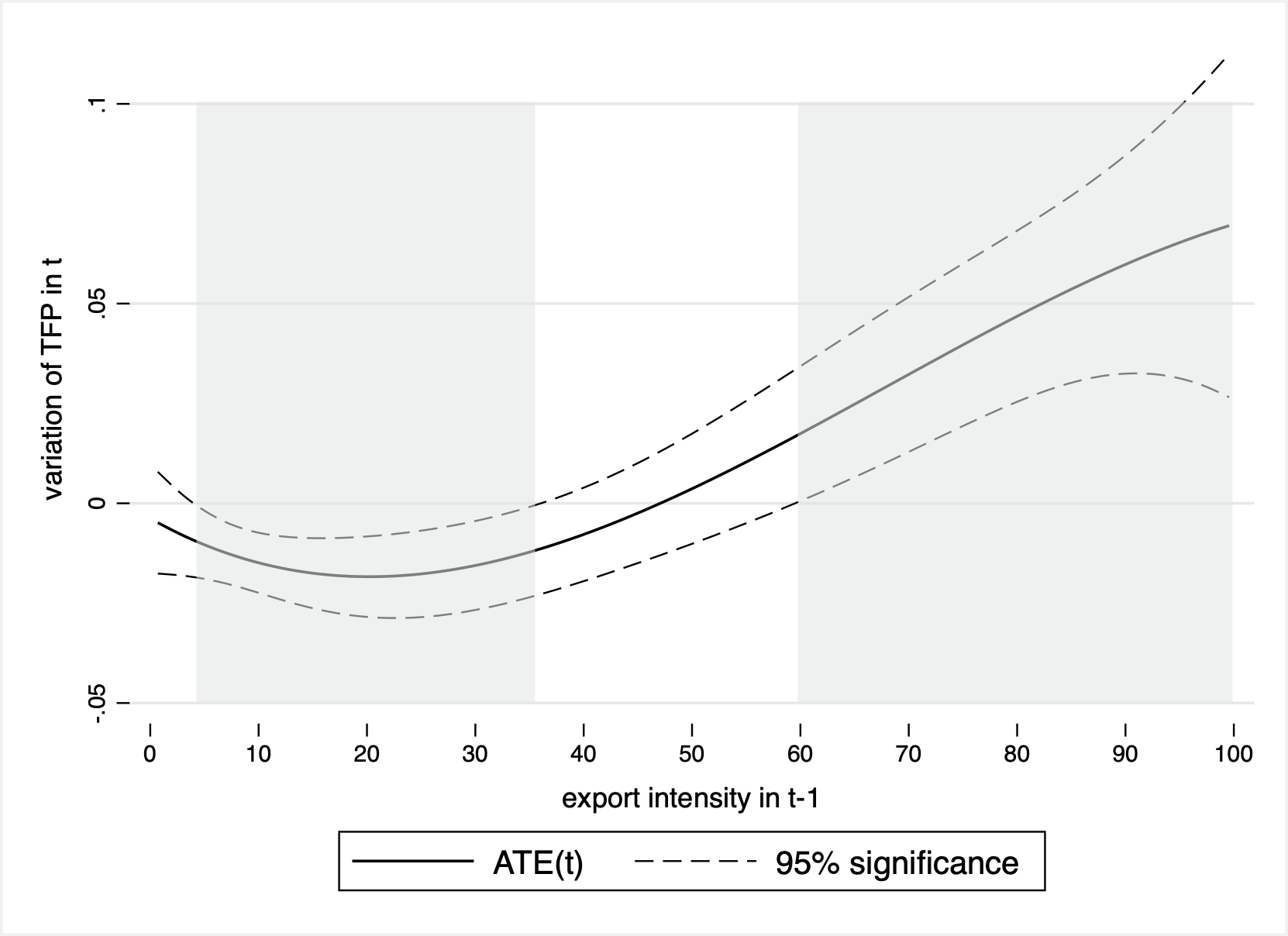}  
        \caption{Polynomial degree 3}
    \end{subfigure}
    \hfill
    \begin{subfigure}[(c)]{0.49\textwidth}
        \includegraphics[width=\textwidth]{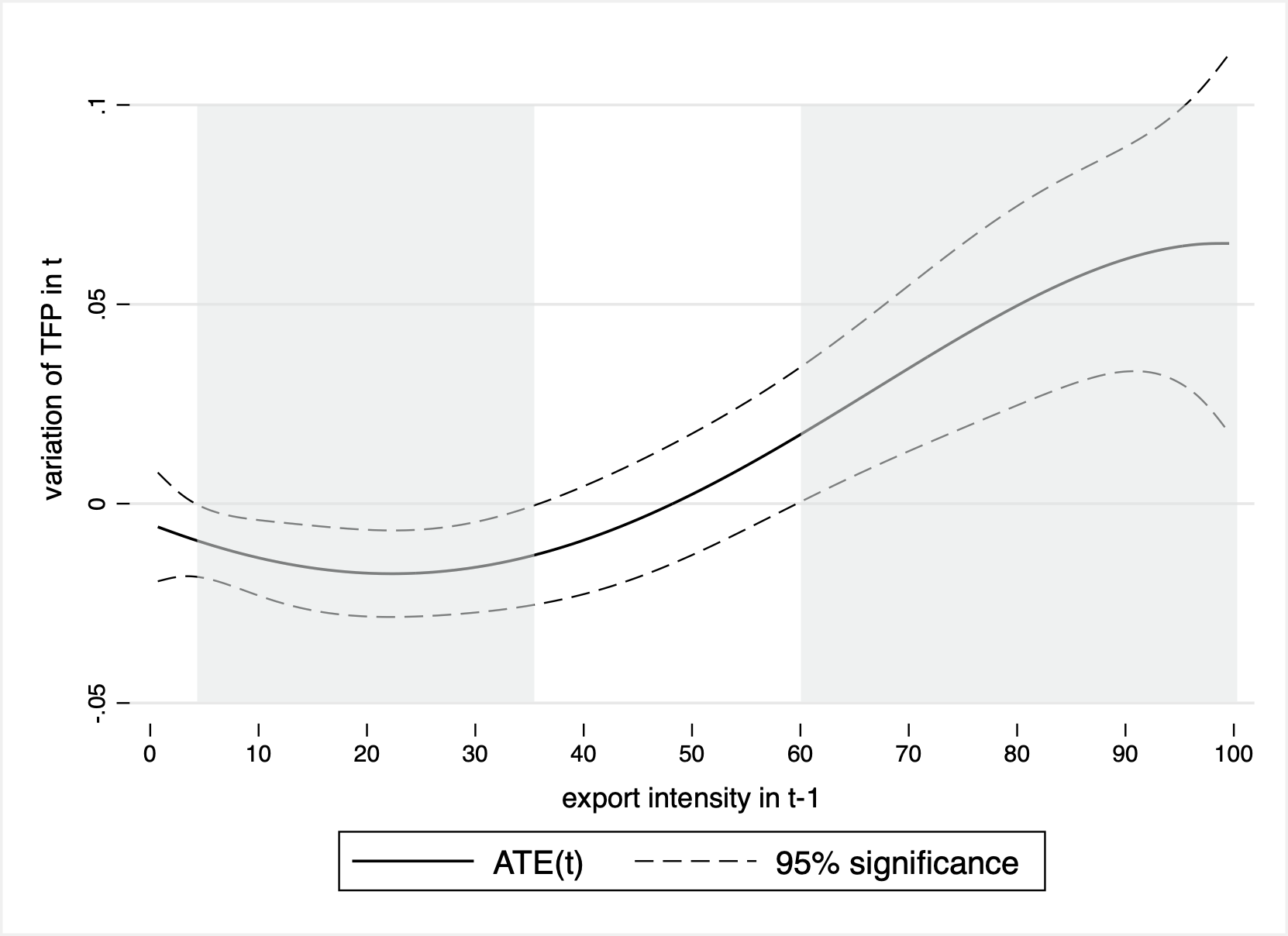}  
        \caption{Polynomial degree 4}
    \end{subfigure}
    \\
    \begin{subfigure}[(d)]
    {0.49\textwidth}
    \includegraphics[width=\textwidth]{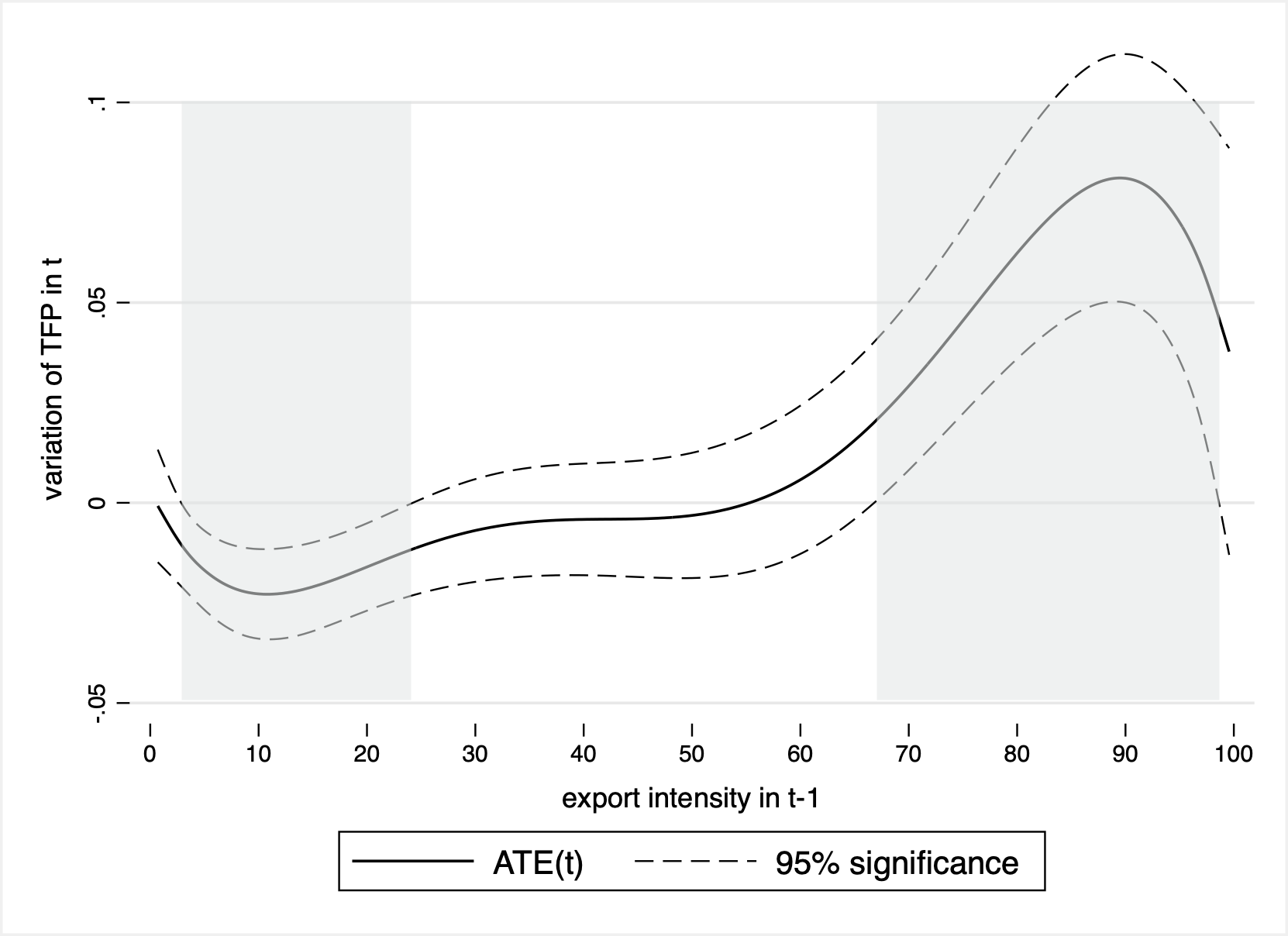} 
    \caption{Polynomial degree 5}
    \end{subfigure}
    \begin{tablenotes}
    \singlespacing \footnotesize
    \item \textit{Note:} In figure we report the estimated dose-response functions when exploring alternative polynomial degrees from 1 (linear case) in figure (a) up to degree 5 in figure (e). Figure (c) corresponds to our baseline, which assumes $h(t)$ to be a polynomial of degree 3.
\end{tablenotes}
\end{figure}

\begin{table}[H]
    \centering
        \caption{Regression models for TFP when we restrict to regions at the border, or including year-industry fixed effects}
    \label{app_tab:robustness_border_ys_FE}
    \resizebox{.5\textwidth}{!}{
    \begin{tabular}{lcc}
    \toprule
    &TFP&TFP\\
&(1)&(2)\\
    \midrule
$D_1$& -0.00148 &-0.00150 \\
& (0.00121)& (0.000899)\\
[1em]
$D_2$&   0.0000300&0.0000423\\
& (0.0000325)& (0.0000243)\\
[1em]
$D_3$&  -9.03e-08& -2.28e-07\\
&  (2.35e-07)  &(1.76e-07)\\

\midrule
Polynomial degree h(t)&3&3\\
firm FE&YES&YES\\
year FE&YES&YES\\
industry-year FE& NO &YES\\
Exporters&non-temporary&non-temporary\\
Regions &at the border&all\\
\midrule
(N)&39,365         &       39,365         \\
\bottomrule

    \end{tabular}}
\begin{tablenotes}
    \singlespacing \footnotesize
    \item \textit{Note:} In this table we report in Column (1) the results of the model estimated considering only regions at the border with other countries.  In Column (2) we report the results of the model estimated when including industry-year fixed effects, to account for industry demand shocks.
\end{tablenotes}
\end{table}

\begin{figure}[H]
\centering
    \caption{Exporters' transition matrix}
    \label{fig: exp_dynamics}
    \includegraphics[width=0.7\textwidth]{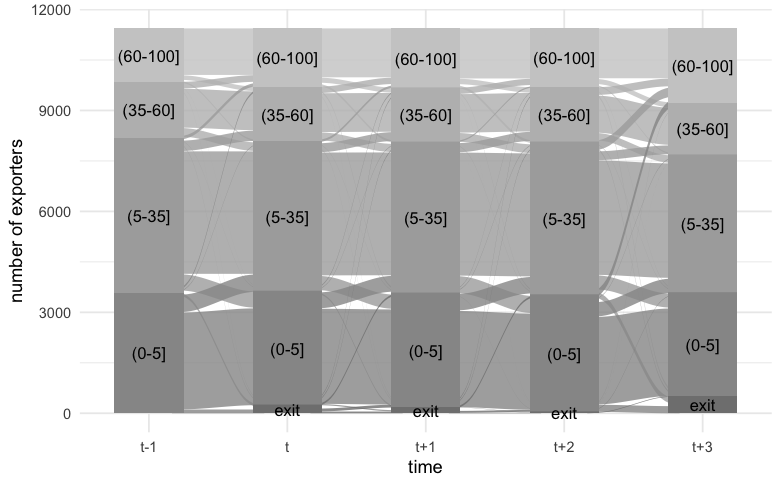}
    
\begin{tablenotes}
    \singlespacing \footnotesize
    \item \textit{Note:} The figure illustrates the exporters' transition in intensity classes over time. Export intensity classes are defined by statistically significant segments obtained from the dose-response function in Figure \ref{fig:drf_TFP}. Flows are directed rightwards from one category to the other, and they visualize the share of exporters changing classes.
\end{tablenotes}
\end{figure}

\newpage

\begin{table}[htpb]
    \centering
        \caption{Balancing properties in the first quintile of the propensity scores distribution}
        \resizebox{\textwidth}{!}{
    \begin{tabular}{l|cccc}
    \toprule
        Variable &  Non-exporting firms & Exporting firms & Mean Difference & p-value\\
        \midrule
Capital Adequacy ratio	&-0.051371&	0.0416095&	0.0929805&	0.1469018\\
Capital Intensity	&0.1780421	&0.1457349	&-0.0323072	&0.4350644\\
Corporate Control&	0.8093995	&0.8192787	&0.0098792	&0.6245197\\
Debtors	&-0.2072747&	-0.089369	&0.1179057*&	0.029857\\
Financial Expenditure	&-0.1394159	&-0.1358956	&0.0035203	&0.9342009\\
Financial Sustainability	&-0.0432416	&-0.0475994	&-0.0043577	&0.7057377\\
Intangible Fixed Assets	&-0.136352&	-0.119541	&0.016811	&0.7694924\\
Inward FDI&	0.6396867	&0.6609465&	0.0212598&	0.3917653\\
Long-term Debt	&-0.0225445	&0.0456641&	0.0682086&	0.1793343\\
Net Profit Margin&	0.3159559&	0.2602014&	-0.0557545&	0.1569565\\
Number of employees&	-0.3268267&	-0.252921	&0.0739058&	0.1293173\\
Operating Revenues Turnover	&-0.0547723	&-0.0299562	&0.0248161	&0.5708422\\
Outward FDI	&0.0261097&	0.0158198&	-0.0102898	&0.1211848\\
P/L after tax&	-0.41663	&-0.3693758	&0.0472542	&0.4381181\\
Productive Capacity	&-0.0823095	&-0.0859178&	-0.0036083&	0.925292\\
Size-age&	-0.3718256	&-0.3686495&	0.0031761&	0.9496028\\
Stock of patents&	0.4222829	&0.3822768	&-0.0400061&	0.5730096\\
Tangible Fixed Assets&	0.019977	&0.0880534	&0.0680765&	0.1204523\\
Total Factor Productivity (ACF)	&-0.1274855&	-0.0908146	&0.0366709&	0.4546481\\
Wage	&0.1842079&	0.0414417	&-0.1427662*&	0.0135933\\
Working Capital&	-0.2837587&	-0.1637572&	0.1200015&	0.0660854\\

\bottomrule
    \end{tabular}
    }
    \label{tab:quintile1}
    \begin{tablenotes}
        \footnotesize \singlespacing 
        \item \textit{Note:} The continuous variables presented here were first log-transformed to reduce distribution skewness, then standardized. Corporate Control, and Outward and Inward FDI are, instead, dummies.
    \end{tablenotes}
\end{table}
\begin{table}[H]
    \centering
        \caption{Balancing properties in the second quintile of the propensity scores distribution}
        \resizebox{\textwidth}{!}{
    \begin{tabular}{l|cccc}
    \toprule
        Variable &  Non-exporting firms & Exporting firms & Mean Difference & p-value\\
        \midrule
Capital Adequacy ratio&0.0030156&-0.0126342&-0.0156499&0.7483503\\
Capital Intensity&0.0147632&0.060906&0.0461428&0.3010347\\
Corporate Control&0.812709&0.8337752&0.0210662&0.3382309\\
Debtors&0.0980864&0.1144996&0.0164132&0.6867808\\
Financial Expenditure&0.0386211&0.11309&0.0744689*&0.0302593\\
Financial Sustainability&-0.0489636&-0.0406799&0.0082838&0.2657352\\
Intangible Fixed Assets&0.0814569&0.0680736&-0.0133832&0.8113292\\
Inward FDI&0.6655518&0.654825&-0.0107268&0.7019553\\
Long-term Debt&0.2437953&0.1593179&-0.0844774&0.1271094\\
Net Profit Margin&0.1410433&0.1191244&-0.0219189&0.5692211\\
Number of employees&-0.1091053&-0.1138495&-0.0047442&0.9234344\\
Operating Revenues Turnover&0.0541782&0.0862548&0.0320766&0.4824737\\
Outward FDI&0.0267559&0.0221368&-0.0046191&0.5959178\\
P/L after tax&-0.0608107&-0.0826253&-0.0218146&0.726598\\
Productive Capacity&-0.0358056&-0.0354333&0.0003723&0.9932139\\
Size-age&-0.0427069&-0.0794048&-0.0366979&0.4722575\\
Stock of patents&0.0930338&0.1307758&0.0377419&0.572309\\
Tangible Fixed Assets&0.1902857&0.2175258&0.0272401&0.4947495\\
Total Factor Productivity (ACF)&-0.1056417&-0.0774978&0.0281439&0.5997008\\
Wage&-0.2329184&-0.0435467&0.1893717***&0.0002419\\
Working Capital&0.0760236&0.041814&-0.0342096&0.5465286\\
\bottomrule
    \end{tabular}
    }
    \label{tab:quintile2}
    \begin{tablenotes}
        \footnotesize \singlespacing 
        \item \textit{Note:} The continuous variables presented here were first log-transformed to reduce distribution skewness, then standardized. Corporate Control, and Outward and Inward FDI are, instead, dummies.
    \end{tablenotes}
\end{table}

\begin{table}[H]
    \centering
        \caption{Balancing properties in the third quintile of the propensity scores distribution}
        \resizebox{\textwidth}{!}{
    \begin{tabular}{l|cccc}
    \toprule
        Variable &  Non-exporting firms & Exporting firms & Mean Difference & p-value\\
        \midrule
Capital Adequacy ratio&-0.0554493&-0.0625667&-0.0071174&0.8838497\\
Capital Intensity&0.0047848&0.0109099&0.0061251&0.8987467\\
Corporate Control&0.875&0.8320137&-0.0429863&0.0653616\\
Debtors&0.2731609&0.2319929&-0.0411681&0.3388023\\
Financial Expenditure&0.2311292&0.2517914&0.0206622&0.5503151\\
Financial Sustainability&-0.0425251&-0.0341136&0.0084115&0.2411463\\
Intangible Fixed Assets&0.165318&0.1903906&0.0250725&0.6407163\\
Inward FDI&0.6666667&0.6541304&-0.0125363&0.673739\\
Long-term Debt&0.0067855&0.1363072&0.1295217*&0.0341994\\
Net Profit Margin&0.0159298&0.0193602&0.0034304&0.9298058\\
Number of employees&0.0399987&-0.0015696&-0.0415683&0.4320493\\
Operating Revenues Turnover&0.2467269&0.2050813&-0.0416457&0.4018219\\
Outward FDI&0.0340909&0.0436791&0.0095882&0.4521335\\
P/L after tax&0.0780331&0.0996515&0.0216184&0.7172298\\
Productive Capacity&0.0542541&-0.0135061&-0.0677602&0.1788279\\
Size-age&0.1098787&0.1269506&0.0170718&0.7463125\\
Stock of patents&-0.0256307&0.0190671&0.0446978&0.4851805\\
Tangible Fixed Assets&0.2938978&0.2641855&-0.0297123&0.4859344\\
Total Factor Productivity (ACF)&-0.0366352&-0.0681151&-0.0314799&0.5865703\\
Wage&-0.0792569&-0.0748793&0.0043777&0.9327033\\
Working Capital&0.1167995&0.1150086&-0.0017909&0.9743612\\
\bottomrule
    \end{tabular}
    }
    \label{tab:quintile3}
    \begin{tablenotes}
        \footnotesize \singlespacing 
        \item \textit{Note:} The continuous variables presented here were first log-transformed to reduce distribution skewness, then standardized. Corporate Control, and Outward and Inward FDI are, instead, dummies.
    \end{tablenotes}
\end{table}

\begin{table}[H]
    \centering
        \caption{Balancing properties in the fourth quintile of the propensity scores distribution}
        \resizebox{\textwidth}{!}{
    \begin{tabular}{l|cccc}
    \toprule
        Variable &  Non-exporting firms & Exporting firms & Mean Difference & p-value\\
        \midrule
Capital Adequacy ratio&-0.0879348&-0.1200475&-0.0321127&0.5409724\\
Capital Intensity&0.0085157&0.0131288&0.0046131&0.9327372\\
Corporate Control&0.8318965&0.8414137&0.0095172&0.6961041\\
Debtors&0.50124&0.4328062&-0.0684338&0.1424907\\
Financial Expenditure&0.4337766&0.4356301&0.0018535&0.9611796\\
Financial Sustainability&-0.0338964&-0.0172256&0.0166708&0.0860451\\
Intangible Fixed Assets&0.2926806&0.2965659&0.0038853&0.9422323\\
Inward FDI&0.6508621&0.6510314&0.0001693&0.9957476\\
Long-term Debt&0.2393196&0.0620695&-0.1772501*&0.0111436\\
Net Profit Margin&-0.1885716&-0.0890155&0.099556*&0.0146271\\
Number of employees&0.2489323&0.188111&-0.0608213&0.304282\\
Operating Revenues Turnover&0.4850428&0.4194674&-0.0655753&0.2441202\\
Outward FDI&0.0775862&0.0867166&0.0091304&0.6259584\\
P/L after tax&0.284603&0.1887881&-0.0958149&0.117392\\
Productive Capacity&-0.0315398&0.0636341&0.095174&0.1064175\\
Size-age&0.3676726&0.4155427&0.0478701&0.4350889\\
Stock of patents&-0.075775&-0.0027727&0.0730023&0.2827041\\
Tangible Fixed Assets&0.4099297&0.3763113&-0.0336183&0.4942812\\
Total Factor Productivity (ACF)&0.03697&0.0304026&-0.0065673&0.9231693\\
Wage&-0.0977594&-0.0777936&0.0199659&0.721956\\
Working Capital&0.2915098&0.1796408&-0.111869&0.0584549\\
\bottomrule
    \end{tabular}
    }
    \label{tab:quintile4}
    \begin{tablenotes}
        \footnotesize \singlespacing 
        \item \textit{Note:} The continuous variables presented here were first log-transformed to reduce distribution skewness, then standardized. Corporate Control, and Outward and Inward FDI are, instead, dummies.
    \end{tablenotes}
\end{table}

\begin{table}[H]
    \centering
        \caption{Balancing properties in the fifth quintile of the propensity scores distribution}
        \resizebox{\textwidth}{!}{
    \begin{tabular}{l|cccc}
    \toprule
        Variable &  Non-exporting firms & Exporting firms & Mean Difference & p-value\\
        \midrule
Capital Adequacy ratio&-0.0155421&-0.0864276&-0.0708855&0.4203504\\
Capital Intensity&0.1864643&0.1399031&-0.0465612&0.524238\\
Corporate Control&0.9130435&0.9106933&-0.0023502&0.9120299\\
Debtors&0.868438&0.8617555&-0.0066825&0.9078007\\
Financial Expenditure&0.8333922&0.8416449&0.0082528&0.8593485\\
Financial Sustainability&0.0793593&0.0466842&-0.0326752&0.2962866\\
Intangible Fixed Assets&0.5649889&0.5007252&-0.0642637&0.2767896\\
Inward FDI&0.6684783&0.6577882&-0.01069&0.7626287\\
Long-term Debt&-0.1938018&-0.0960866&0.0977152&0.2584741\\
Net Profit Margin&-0.2582484&-0.2390168&0.0192316&0.7204699\\
Number of employees&0.6646712&0.6515535&-0.0131177&0.8602134\\
Operating Revenues Turnover&0.9107945&0.9309824&0.0201879&0.7722595\\
Outward FDI&0.3423913&0.3467816&0.0043903&0.9016028\\
P/L after tax&0.2679023&0.3010762&0.0331739&0.6312838\\
Productive Capacity&0.1882188&0.1671936&-0.0210252&0.8043682\\
Size-age&1.005051&0.9596214&-0.0454294&0.5980579\\
Stock of patents&0.2695403&0.1074307&-0.1621095&0.066014\\
Tangible Fixed Assets&0.6741868&0.6539947&-0.0201921&0.7519363\\
Total Factor Productivity (ACF)&0.2971215&0.2121911&-0.0849304&0.3204345\\
Wage&0.1837182&0.1202443&-0.0634739&0.3887388\\
Working Capital&0.3049121&0.3136751&0.008763&0.8975379\\
\bottomrule
    \end{tabular}
    }
    \label{tab:quintile5}
    \begin{tablenotes}
       \item  \footnotesize \singlespacing 
       \item \textit{Note:} The continuous variables presented here were first log-transformed to reduce distribution skewness, then standardized. Corporate Control, and Outward and Inward FDI are, instead, dummies.
    \end{tablenotes}
\end{table}

\end{document}